\documentclass[journal]{IEEEtran}
\usepackage{graphicx}
\usepackage{color}
\usepackage{placeins}
\usepackage{float}
\usepackage{tabularx,colortbl}
\usepackage{amssymb}
\usepackage{amsthm}
\usepackage{cite}
\usepackage{amsmath}
\usepackage{makecell}
\usepackage{subfigure}

\usepackage{bm}
\usepackage{cleveref}
\usepackage{multicol}       % table
\usepackage{multirow}       % table
\usepackage{array}          % table
\usepackage{colortbl}
\usepackage{makecell}
\definecolor{crimson}{RGB}{192,0,0}         % color crimson
\definecolor{navy}{RGB}{47,85,151}         % color crimson

\makeatletter                               % algorithm
\newif\if@restonecol
\makeatother

\makeatletter
\newif\if@restonecol
\makeatother

\usepackage[linesnumbered,ruled,vlined]{algorithm2e}%[ruled,vlined]{
\usepackage{algpseudocode}
\usepackage{amsmath}
\renewcommand{\arraystretch}{1.5} %
\usepackage{stfloats}
\usepackage{color}
\usepackage{placeins}
\usepackage{tabularx,colortbl}
\usepackage{amssymb}

\usepackage{listings}                       % Matlab code
\usepackage[framed,numbered,autolinebreaks,useliterate]{mcode}
\theoremstyle{plain}
\newtheorem{thm}{Theorem}

\newtheorem{coro}{Corollary}

\theoremstyle{plain}
\newtheorem{rem}{Remark}

\IEEEoverridecommandlockouts

\begin{document}

%----------------------------title&author&thanks----------------------------

\title{Analytical Framework for Effective Degrees of Freedom in Near-Field XL-MIMO
\thanks{Z. Wang, J. Zhang, W. Yi, and B. Ai are with the State Key Laboratory of Advanced Rail Autonomous Operation, and also with the School of Electronics and Information Engineering, Beijing Jiaotong University, Beijing 100044, P. R. (e-mail: \{zhewang\_77, jiayizhang, 23125012, boai\}@bjtu.edu.cn);}
\thanks{H. Xiao is with ZTE Corporation, State Key Laboratory of Mobile Network and Mobile Multimedia Technology (e-mail: xiao.huahua@zte.com.cn);}
\thanks{H. Du is with the Department of Electrical and Electronic Engineering, University of Hong Kong, Pok Fu Lam, Hong Kong SAR, China (e-mail: duhy@eee.hku.hk);}
\thanks{D. Niyato is with the College of Computing \& Data Science, Nanyang Technological University, Singapore 639798 (e-mail: dniyato@ntu.edu.sg);}
\thanks{D. W. K. Ng is with the School of Electrical Engineering and Telecommunications, University of New South Wales, Sydney, NSW 2052, Australia (e-mail: w.k.ng@unsw.edu.au).}}
\author{Zhe Wang, Jiayi Zhang,~\IEEEmembership{Senior Member,~IEEE,} Wenhui Yi, Huahua Xiao, Hongyang Du, \\Dusit Niyato,~\IEEEmembership{Fellow,~IEEE}, Bo Ai,~\IEEEmembership{Fellow,~IEEE}, and Derrick Wing Kwan Ng,~\IEEEmembership{Fellow,~IEEE}}
\maketitle
\maketitle

%----------------------------abstract----------------------------

\begin{abstract}
Extremely large-scale multiple-input-multiple-output (XL-MIMO) is an emerging transceiver technology for enabling next-generation communication systems, due to its potential for substantial enhancement in both the spectral efficiency and spatial resolution. However, the achievable performance limits of various promising XL-MIMO configurations have yet to be fully evaluated, compared, and discussed. In this paper, we develop an effective degrees of freedom (EDoF) performance analysis framework specifically tailored for near-field XL-MIMO systems. We explore five representative distinct XL-MIMO hardware designs, including uniform planar array (UPA)-based with infinitely thin dipoles, two-dimensional (2D) continuous aperture (CAP) plane-based, UPA-based with patch antennas, uniform linear array (ULA)-based, and one-dimensional (1D) CAP line segment-based XL-MIMO systems. Our analysis encompasses two near-field channel models: the scalar and dyadic Green's function-based channel models. More importantly, when applying the scalar Green's function-based channel, we derive EDoF expressions in the closed-form, characterizing the impacts of the physical size of the transceiver, the transmitting distance, and the carrier frequency. In our numerical results, we evaluate and compare the EDoF performance across all examined XL-MIMO designs, confirming the accuracy of our proposed closed-form expressions. Furthermore, we observe that with an increasing number of antennas, the EDoF performance for both UPA-based and ULA-based systems approaches that of 2D CAP plane and 1D CAP line segment-based systems, respectively. Moreover, we unveil that the EDoF performance for near-field XL-MIMO systems is predominantly determined by the array aperture size rather than the sheer number of antennas.
\end{abstract}
%----------------------------keywords----------------------------
\begin{IEEEkeywords}
XL-MIMO, effective degrees of freedom, near-field communication, polarization effect.
\end{IEEEkeywords}

%\newpage
\IEEEpeerreviewmaketitle

\section{Introduction}\label{intro}
The rapid development of wireless communication has sparked considerable research interest in beyond fifth-generation (B5G) and sixth-generation (6G) wireless communication networks. Compared to existing 5G networks, 6G networks are anticipated to exhibit dramatically enhanced communication capabilities, such as a 100-fold increase in peak data rate (reaching the Tb/s level), a tenfold reduction in latency, and an end-to-end reliability requirement of $99.99999\%$ \cite{you2021towards,9390169,9113273}. To attain the aforementioned communication capabilities and satisfy various stringent communication demands, a variety of promising technologies have been widely studied, such as extremely large-scale multiple-input multiple-output (XL-MIMO) \cite{ZheSurvey,2023arXiv231011044L,9903389,2022arXiv221201257G}, reconfigurable intelligent surfaces (RIS) \cite{8811733,enyusurvey,9808307,sang2023multi}, and artificial intelligence (AI)-aided intelligent networks \cite{du2023ai,10051719}. Among these technologies, XL-MIMO is regarded as an evolutionary paradigm in MIMO technology, which embraces an extraordinarily large number of antennas and supremely large array aperture, thereby facilitating effective communications \cite{ZheSurvey,2023arXiv231011044L}. 

The fundamental concept of XL-MIMO involves utilizing arrays with extremely large aperture, achieved by deploying an extremely large number of antennas, e.g. thousands of antennas \cite{ZheSurvey}. In practice, XL-MIMO can be classified into three main categories: uniform linear array (ULA)-based, uniform planar array (UPA)-based, and continuous aperture (CAP)-based designs. Specifically, the ULA-based design exhibits one-dimensional (1D) array characteristics similar to those of conventional massive MIMO systems \cite{[162],sanguinetti2019toward,OBETrans}, but with a significantly higher number of antennas, such as 512 antennas or even thousands. As for the UPA-based design, typically involving a rectangular or square plane array with thousands of antennas, exhibits two-dimensional (2D) array characteristics. Two prevalent forms of the antenna element are commonly considered in the literature: sizeless infinitely thin dipoles \cite{[15]} and patch antennas with a certain physical size \cite{[97]}. This 2D planar array has been widely studied in full-dimension MIMO technology, which is viewed as an vital enabler of late fourth-generation (4G)/early 5G wireless networks. This 2D planar array has been widely studied in full-dimension MIMO technology, which is considered a vital enabler of late fourth-generation (4G) and early 5G wireless networks \cite{nam2013full,kim2014full}. In full-dimension MIMO, antennas are placed in a 2D grid at the base stations. Moreover, the CAP-based design represents a novel scheme characterized by an approximately continuous array aperture.
This design, facilitated by meta-materials, achieves an extremely dense deployment of an approximately infinite number of antennas \cite{[70]}. Different from other XL-MIMO designs, the CAP-based XL-MIMO design leverages spatially-continuous electromagnetic (EM) characteristics that requires integral-based signal processing and performance analysis techniques. These general XL-MIMO configurations have sparked extensive research efforts. Note that one distinguishing characteristic in XL-MIMO systems is the consideration of near-field characteristics, aided by spherical waves, in contrast to the planar wave EM characteristics considered in conventional far-field based massive MIMO systems \cite{ZheSurvey,2024arXiv240105900L,10220205}. In fact, the EM spherical wave characteristics can be described by the Green's function-based channel model \cite{arnoldus2001representation,ZheSurvey}. Based on the spherical wave characteristics, numerous studies have focused on near-field XL-MIMO systems: near-field channel modeling \cite{[41],2022arXiv220903082R,[34]}, near-field channel estimation \cite{[6],[35],chen-jsac3}, near-field beamforming \cite{10243590,[101]}, and near-field resource allocation \cite{2023arXiv230616206Y,Zhilongmag}.

To highlight the advantages of near-field XL-MIMO systems, it is imperative to study their performance limits. Degrees of freedom (DoF) is recognized as an important performance metric for wireless communication \cite{franceschetti2017wave}. In general, the number of DoF can be regarded as the dimension of the signals' space available for communication \cite{franceschetti2017wave}. Mathematically, the rigorous definition of the number of DoF, up to an accuracy $\epsilon >0$, was defined in \cite{franceschetti2017wave} as the index of the smallest ordered eigenvalue that falls below an arbitrarily small threshold $\epsilon$. Following this definitive standard, many works have endeavoured to explore the DoF performance for MIMO systems \cite{poon2005degrees,franceschetti2015landau,pizzo2022landau}. In MIMO systems, the number of DoF can be regarded as the number of independent signal dimensions over a wireless channel, which can be applied for transmitting information. As observed in \cite[Fig. 2]{franceschetti2015landau}, the eigenvalues of the MIMO spatial channel exhibit a step function-like behavior, thus an accuracy level $\epsilon >0$ can be also included to quantify the number of eigenvalues above this accuracy level. More specifically, the insightful Landau's eigenvalue theorem was applied in \cite{franceschetti2015landau} and \cite{pizzo2022landau} to derive the number of eigenvalues above an accuracy level $\epsilon >0$.

It is worth noting that these works focusing on the DoF analysis for MIMO systems \cite{poon2005degrees,franceschetti2015landau,pizzo2022landau} pose very rigorous derivations of the DoF. However, these results rely on demanding mathematical and sampling backgrounds, and are not easy to exploit for tractable performance analysis of MIMO systems. Indeed, it is promising to provide handy and tractable DoF performance analysis framework for MIMO systems. In near-field XL-MIMO systems, relying on the extremely large array aperture, extensive simulation results demonstrate that the characteristic of the ordered eigenvalues of the channel can be divided into two stages. In the first stage, the singular values maintain approximately constant and then, at one particular singular value, sharply degrade approaching to zero as illustrated in \cite[Fig. 7]{2024arXiv240105900L}. This phenomenon is not isolated but corresponds to a similar step function-like phenomenon as in \cite{poon2005degrees,franceschetti2015landau,pizzo2022landau}, and the decay of singular values becomes very sharp due to the extremely large number of antennas. Exploiting this characteristic, the index of the singular value, where sharp decay begins, is defined as effective DoF (EDoF) in extensive works on near-field XL-MIMO systems \cite{ouyangEDoF,[65],[29],2023arXiv230406141X,EDOFTVT}. Relying on the characteristic of these singular values, an approximate estimation of the EDoF, as will be shown in \eqref{EDoF_Scalar_Dis} below, is widely studied due to its intuitive form and tractable potential. In summary, \emph{the EDoF concept can be viewed as an approximate mathematical solution for determining the number of dominant singular values or sub-channels in these systems.} More importantly, this concept not only corresponds to the rigorous definitions in classical related works  \cite{franceschetti2017wave,poon2005degrees,franceschetti2015landau,pizzo2022landau} but also offers an intuitive and tractable formulation.

\renewcommand\arraystretch{1.5}
\begin{table*}[t!]
  \centering
  \fontsize{9}{11}\selectfont
  \caption{Comparison of relevant papers with our paper.}
  \label{comparison}
    \begin{tabular}{|p{1.2cm}<{\centering}|p{1.8cm}<{\centering}|p{1.8cm}<{\centering}|p{1cm}<{\centering}|p{1.2cm}<{\centering}|p{1cm}<{\centering}|p{1.2cm}<{\centering}|p{2cm}<{\centering}|p{2cm}<{\centering}|}
    \hline
      \multirow{2}{*}{\textbf{Ref.}} & \multicolumn{2}{c|}{\textbf{Channel Model}} & \multicolumn{2}{c|}{\textbf{1D Schemes}} & \multicolumn{2}{c|}{\textbf{2D Schemes}} & \multicolumn{2}{c|}{\textbf{Closed-form Results}} \cr\cline{2-9}
      & Scalar Green & Dyadic Green & ULA & 1D CAP & UPA & 2D CAP & 1D Schemes & 2D Schemes \cr\hline
      \cite{2023arXiv230406141X} &  $\surd$  &  $\times$ &  $\surd$ &  $\surd$ & $\times$ & $\times$ & $\surd$ & $\times$\cr\hline
      \cite{[65]} & $\surd$ & $\surd$ & $\times$ & $\times$ & $\surd$ & $\times$ & $\times$ & $\times$ \cr\hline
      \cite{[29]} & $\surd$  &  $\times$ & $\surd$ & $\surd$ & $\times$ & $\times$ & $\surd$  & $\times$ \cr\hline
      \cite{ouyangEDoF} & $\surd$ &  $\times$ & $\surd$ & $\surd$ & $\times$ & $\times$ & $\times$ & $\times$ \cr\hline
      \textbf{Proposed} & $\surd$ & $\surd$ & $\surd$ & $\surd$ & $\surd$ & $\surd$ & $\surd$ & $\surd$ \cr\hline
    \end{tabular}
\end{table*}

Based on this concept, the EDoF performance analysis for different XL-MIMO designs has been implemented. For 1D XL-MIMO designs, the authors in \cite{[29]} proposed an EDoF analysis framework for ULA-based and 1D CAP line segment-based XL-MIMO systems, utilizing the scalar Green's function-based channel. Building upon this framework, the authors in \cite{2023arXiv230406141X} derived closed-form EDoF expressions for both the ULA-based and 1D CAP line segment-based XL-MIMO systems over the scalar Green's function-based channel. Subsequently, they compared the EDoF performance between these two XL-MIMO designs. As for the 2D XL-MIMO design, the authors in \cite{[65]} investigated the EDoF performance for UPA-based XL-MIMO systems with infinitely thin dipoles, exploring both scalar and dyadic Green's function-based channels. The findings in \cite{[65]} revealed that employing the dyadic Green's function-based channel could potentially discover more inherent EDoF compared with the scalar Green's function-based channel. This improvement can be attributed to the inclusion of triple polarization effects in the dyadic channel model. Further expanding the research scope, the authors in \cite{EDOFTVT} proposed an EDoF performance analysis framework for 2D CAP plane-based XL-MIMO systems. This framework leveraged the asymptotic analysis technique and the EDoF performance was compared between 2D CAP plane-based XL-MIMO systems and UPA-based XL-MIMO systems with infinitely thin dipoles.

The existing research on EDoF performance analysis has been conducted separately, focusing solely on one or two specific XL-MIMO hardware designs. However, it is vital to comprehensively exploit and compare the EDoF performance for all promising XL-MIMO designs to provide insightful guidelines for evaluating EDoF performance for XL-MIMO \cite{ZheSurvey}. Additionally, in practice, BSs are typically equipped with plane-based antenna arrays, which are flexible to deploy and effective to utilize the full-dimension spatial resources to enhance system performance. However, taking the CAP-based XL-MIMO system as an example, compared with 1D XL-MIMO systems, analyzing the EDoF performance for 2D XL-MIMO systems poses significant computational complexity due to the involvement of solving octuple integrals. Therefore, the derivation of EDoF closed-form expressions becomes essential to facilitate efficient performance analysis of XL-MIMO systems. Note that existing EDoF closed-form expressions tailored for 1D XL-MIMO systems involve the computation of quadruple integrals \cite{2023arXiv230406141X}. In contrast, deriving similar expressions for 2D XL-MIMO systems poses significantly greater computational hurdles, primarily due to the complexity introduced by octuple integrals. Thus, a closed-form analytical framework for the EDoF performance of 2D XL-MIMO systems is highly anticipated. Motivated by the above observations, we study the EDoF performance for near-field XL-MIMO systems in this paper. The comparisons of relevant papers, working on the EDoF analysis of near-field XL-MIMO systems, with this paper are summarized in Table~\ref{comparison}. The main contributions are given as follows.

\begin{itemize}
\item We investigate the EDoF performance analysis framework for evaluating the performance of five representative XL-MIMO designs: UPA-based with infinitely thin dipoles, 2D CAP plane-based, UPA-based with patch antennas, ULA-based, and 1D CAP line segment-based XL-MIMO systems over two channel models, the scalar and dyadic Green's function-based channel models.
\item For the scalar Green's function-based channel, we derive novel closed-form EDoF performance expressions for XL-MIMO systems. More importantly, extensive simulations are performed to validate the accuracy of our derived closed-form results and highlight important system design insights.
\item We comprehensively explore and compare the EDoF performance for all studied XL-MIMO designs. Our observations reveal that the EDoF performance for discrete XL-MIMO designs, i.e., UPA and ULA-based XL-MIMO designs, approaches that of respective continuous XL-MIMO designs as the number of antennas increases, i.e., 2D CAP plane and 1D CAP line segment-based XL-MIMO designs. And the EDoF performance can be improved by incorporating multiple polarization and by increasing the physical sizes of the transceiver. More importantly, the EDoF performance primarily depends on the physical size of the array physical size rather than the sheer number of antennas.
\end{itemize}

The rest of this paper is organized as follows. In Section~\ref{point}, we study the EDoF performance for UPA-based XL-MIMO systems with infinitely thin dipoles. Then, Section~\ref{2DCAP} develops an EDoF performance analysis framework for 2D CAP-based XL-MIMO systems. More significantly, novel EDoF closed-form expressions are derived over the scalar Green's function-based channels. In Section~\ref{other}, we study the EDoF performance for UPA-based with patch antennas, ULA-based, and 1D CAP line segment-based XL-MIMO systems. In Section~\ref{num}, we provide extensive simulation results to evaluate, compare, and discuss the EDoF performance for all considered XL-MIMO designs. Finally, a brief conclusion is drawn in Section~\ref{con}.

\textbf{\emph{Notation}}: Let boldface lowercase letters $\mathbf{x}$ and boldface uppercase letters $\mathbf{X}$ denote column vectors and matrices, respectively. $\mathbb{C}$ and $\mathbb{R}$ denote the set of complex and real, respectively. We denote $\mathbf{I}_{n\times n}$ as the $n\times n$ identity matrix. $\nabla _{\mathbf{x}}$ is the first-order partial derivative operator with respect to $\mathbf{x}$. We define $\left( \cdot \right) ^T$, $\left( \cdot \right) ^*$, and $\left( \cdot \right) ^H$ as the transpose, conjugate, and conjugate transpose, respectively. $\mathrm{mod}\left( \cdot ,\cdot \right) $ and $\lfloor \cdot \rfloor $ denote the modulus operation and the truncation operation, respectively. $\mathrm{tr}\left\{ \cdot \right\}$ and $\triangleq$ are the trace operator and the definitions, respectively. Also, we denote $\left\| \cdot \right\|$ and $\left\| \cdot \right\| _{\mathrm{F}}$ as the Euclidean norm and the Frobenious norm, respectively.

\section{UPA-Based XL-MIMO with Infinitely Thin Dipoles}\label{point}
\subsection{System Model}\label{SystemUPA}
\begin{figure}[t]
\centering
\includegraphics[scale=0.35]{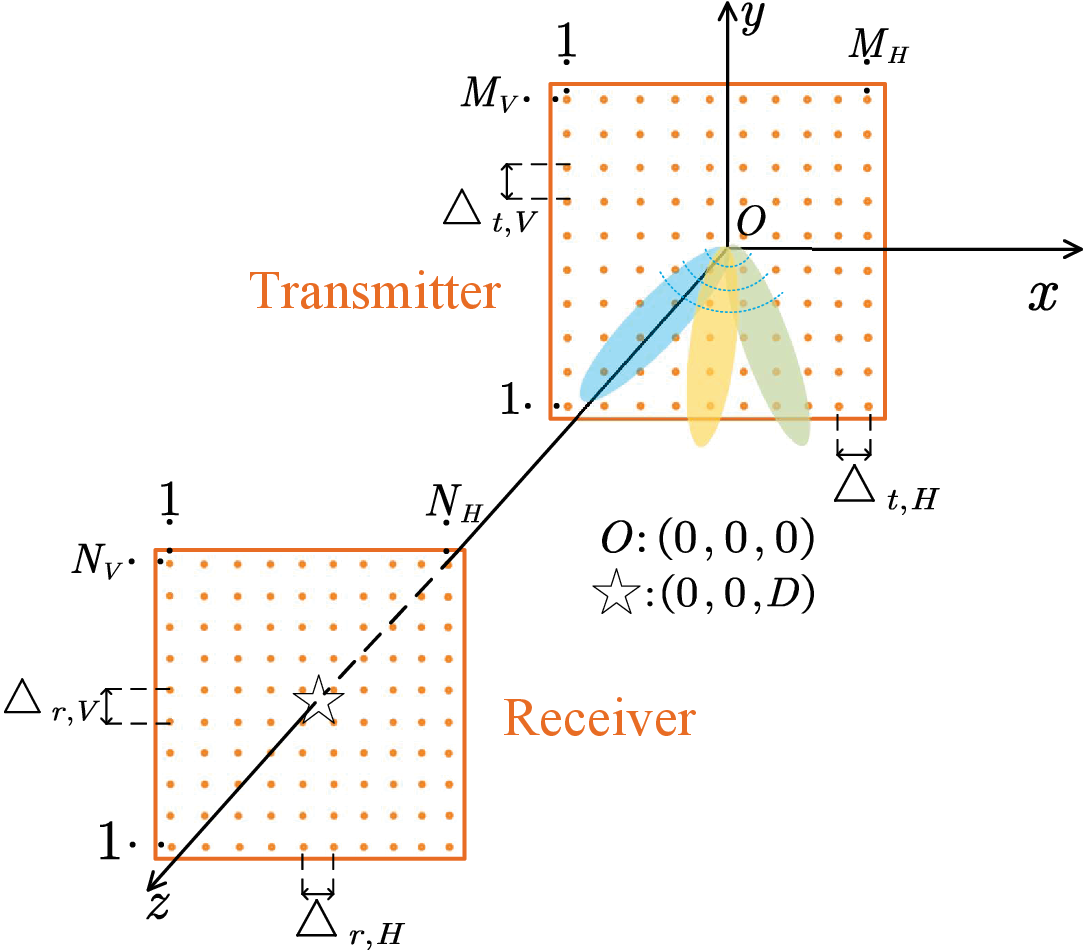}
\caption{A UPA-based XL-MIMO system with each antenna element being a sizeless infinitely thin dipoles.\label{0}}
\vspace{-0.4cm}
\end{figure}
We first study the UPA-based XL-MIMO system with infinitely thin dipoles \cite{[6],[15]}. As illustrated in Fig. \ref{0}, one transmitter and one receiver equipped with UPA-based XL-MIMO with infinitely thin dipoles are considered\footnote{By modelling antennas as infinitely thin dipoles, we can assume that only a sizeless point is impinged to transmit or receive EM waves, which simplifies the analysis. To achieve this, there are two possible approaches. First, we can assume that the physical size of each antenna is extremely small, analogous to infinitely thin dipoles, allowing us to treat each antenna as a sizeless point. Second, we can assume that only a single point on each antenna element, regardless of its physical size, is involved in transmitting or receiving EM waves.}. The transmitter is located in the $x-y$ plane with its center being the origin. The number of transmitting antennas per row, the number of transmitting antennas per column, the horizontal side-length, and the vertical side-length of the transmitter are defined as $M_H$, $M_V$, $L_{t,H}$ and $L_{t,V}$, respectively. Thus, the horizontal and vertical antenna spacing for the transmitter can be denoted as $\Delta _{t,H}=L_{t,H}/M_H$ and $\Delta _{t,V}=L_{t,V}/M_V$, respectively. Also, the total number of transmitting antennas is $M=M_VM_H$. Inspired by \cite{[50]}, let $m\in \left\{ 1,\dots ,M \right\} $ denote the antenna index row-by-row counted from the bottom left of the transmitter. Thus, the horizontal and vertical indices of the $m$-th transmitting antenna can be denoted as $i\left( m \right) =\mathrm{mod}\left( m-1,M_H \right)$ and $j\left( m \right) =\lfloor \left( m-1 \right) /M_H \rfloor$, respectively. We denote the position of the $m$-th transmitting antenna by $\mathbf{r}_{t,m}=[-\frac{L_{t,H}}{2}+i\left( m \right) \Delta _{t,H},-\frac{L_{t,V}}{2}+j\left( m \right) \Delta _{t,V},0]^T\in \mathbb{R} ^3$.

Meanwhile, for the receiver, let $N_H$ and $N_V$ denote the number of receiving antennas per row and per column, respectively. Thus, the receiver is equipped with a total of $N=N_VN_H$ antennas. $L_{r,H}$ and $L_{r,V}$ are the horizontal and vertical side-lengths of the receiver, respectively. Also, the horizontal and vertical antenna spacing for the receiver are $\Delta _{r,H}=L_{r,H}/N_H$ and $\Delta _{r,V}=L_{r,V}/N_V$, respectively. Note that the center of the receiver is located on the $z$ axis with $D$ being the distance between the centers of the transmitter and the receiver. Thus, the position of the $n$-th receiving antenna is defined as $
\mathbf{r}_{r,n}=[-\frac{L_{r,H}}{2}+t\left( n \right) \Delta _{r,H},-\frac{L_{r,V}}{2}+k\left( n \right) \Delta _{r,V},D]^T\in \mathbb{R} ^3$, where $t\left( n \right) =\mathrm{mod}\left( n-1,N_H \right)$ and $k\left( n \right) =\lfloor \left( n-1 \right) /N_H \rfloor$ are the horizontal and vertical indices of the $n$-th receiving antenna, respectively, with $n\in \left\{ 1,\dots ,N \right\} $.

\subsection{Channel Model}
In this paper, we investigate two types of Green's function-based channel models: scalar Green's function-based and dyadic Green's function-based channel models \cite{ZheSurvey,[65]}, which are efficient to describe near-field EM wave characteristics\footnote{Due to the extremely large-scale array aperture, the line-of-sight (LoS) component is significantly dominant. Thus, in this paper, we assume that only LoS channel components exist. The analysis can be easily expanded to scenarios that include both LoS and non-line-of-sight (NLoS) channel components by incorporating additional NLoS components in the channel matrices \cite{ouyangEDoF}.}.
\subsubsection{Scalar Green's Function-Based Channel Model}
For this channel model, the channel matrix between the transmitter and the receiver $\mathbf{H}_S\in \mathbb{C} ^{N\times M}$ is generated based on the scalar Green's function. Note that the scalar Green's function between an arbitrary receiving point $\mathbf{r}\in \mathbb{R} ^3$ and an arbitrary transmitting point $\mathbf{s}\in \mathbb{R} ^3$ can be denoted as \cite{[65],2023arXiv230406141X}
\begin{equation}\label{Scalar_Green}
G\left( \mathbf{r},\mathbf{s} \right) =\frac{1}{4\pi}\frac{\exp \left( -j\kappa _0\left| \mathbf{r}-\mathbf{s} \right| \right)}{\left| \mathbf{r}-\mathbf{s} \right|}
\end{equation}
with $\kappa _0=2\pi/\lambda$ and $\lambda$ being the wavenumber and the wavelength, respectively. Based on \eqref{Scalar_Green}, the channel between the $n$-th receiving antenna and $m$-th transmitting antenna $\left[ \mathbf{H}_S \right] _{nm}=G_{nm}$ is generated by the scalar Green's function as
\begin{equation}\label{Discrete_Scalar_mn}
G_{nm}=G\left( \mathbf{r}_{r,n},\mathbf{r}_{t,m} \right) =\frac{1}{4\pi}\frac{\exp\mathrm{(}-j\kappa _0\left| \mathbf{r}_{r,n}-\mathbf{r}_{t,m} \right|)}{\left| \mathbf{r}_{r,n}-\mathbf{r}_{t,m} \right|}.
\end{equation}
\subsubsection{Dyadic Green's Function-Based Channel Model}
According to \cite{[65],ZheSurvey}, the polarization effect should be considered in near-field XL-MIMO systems and the introduction of multiple polarization can benefit the system performance. Thus, the dyadic Green's function-based channel model is also investigated, where the multi-polarization effect is modeled. The dyadic Green's function between a receiving point $\mathbf{r}\in \mathbb{R} ^3$ and a transmitting point $\mathbf{s}\in \mathbb{R} ^3$ as $\mathbf{G}\left( \mathbf{r},\mathbf{s} \right) \in \mathbb{C} ^{3\times 3}$ is given in \eqref{Dyadic}, where $\vec{\mathbf{a}}_{rs}=\left( \mathbf{r}-\mathbf{s} \right) /\left| \mathbf{r}-\mathbf{s} \right|\in \mathbb{R} ^3$ is a unit direction vector. It is worth noting that the dyadic Green's function, denoted as $\mathbf{G}\in \mathbb{C} ^{3\times 3}$ as in \eqref{Dyadic}, can be expressed in a matrix form
\addtocounter{equation}{1}
\begin{equation}\label{GPolar}
\mathbf{G}=\left[ \begin{matrix}
	G^{xx}&		G^{xy}&		G^{xz}\\
	G^{yx}&		G^{yy}&		G^{yz}\\
	G^{zx}&		G^{zy}&		G^{zz}\\
\end{matrix} \right],
\end{equation}
where $G^{pq}\in \mathbb{C}$ is the scalar Green's function between the polarization direction $p$ of the receiving point and the polarization direction $q$ of the transmitting point with $p,q\in \left\{ x,y,z \right\}$. Furthermore, $G^{pq}$ can be denoted as $G^{pq}=\eta ( p,q ) G( \mathbf{r},\mathbf{s} ) $, where $\eta ( p,q ) =1-\vec{a}_{rs}^{p}\vec{a}_{rs}^{p}+{( j-3j\vec{a}_{rs}^{p}\vec{a}_{rs}^{p} )}/{\kappa _0| \mathbf{r}-\mathbf{s} |}+{( 3\vec{a}_{rs}^{p}\vec{a}_{rs}^{p}-1 )}/{\kappa _{0}^{2}| \mathbf{r}-\mathbf{s} |^2}$ for $p=q$ and $\eta ( p,q ) =( {3}/{\kappa _{0}^{2}| \mathbf{r}-\mathbf{s} |^2}-{3j}/{\kappa _0| \mathbf{r}-\mathbf{s} |}-1 ) \vec{a}_{rs}^{p}\vec{a}_{rs}^{q}$ for $p\ne q$ with $\vec{a}_{rs}^{p}$ being the component of the polarization direction $p$ in the unit direction vector $\vec{\mathbf{a}}_{rs}=[ \vec{a}_{rs}^{x},\vec{a}_{rs}^{y},\vec{a}_{rs}^{z} ] \in \mathbb{R} ^3$.

According to \eqref{Dyadic}, the dyadic Green's function between the $n$-th receiving antenna and the $m$-th transmitting antenna can be represented as
\begin{equation}\label{Discrete_Dyadic_mn}
\mathbf{G}\left( \mathbf{r}_{r,n},\mathbf{r}_{t,m} \right)\! =\! \left( \mathbf{I}_3+\frac{\nabla _{\mathbf{r}_{r,n}}\nabla _{\mathbf{r}_{r,n}}^{H}}{\kappa _{0}^{2}} \right) G\left( \mathbf{r}_{r,n},\mathbf{r}_{t,m} \right),
\end{equation}
where $G\left( \mathbf{r}_{r,n},\mathbf{r}_{t,m} \right)$ is given in \eqref{Discrete_Scalar_mn}.
Based on \eqref{Dyadic} and \eqref{GPolar}, the dyadic Green's function-based channel matrix between the transmitter and the receiver can be denoted as
\begin{equation}\label{HD_Dis}
\begin{aligned}
\mathbf{H}_D=\left[ \begin{matrix}
	\mathbf{H}^{xx}&		\mathbf{H}^{xy}&		\mathbf{H}^{xz}\\
	\mathbf{H}^{yx}&		\mathbf{H}^{yy}&		\mathbf{H}^{yz}\\
	\mathbf{H}^{zx}&		\mathbf{H}^{zy}&		\mathbf{H}^{zz}\\
\end{matrix} \right]\in \mathbb{C} ^{3N\times 3M} ,
\end{aligned}
\end{equation}
where $\mathbf{H}^{pq}\in \mathbb{C} ^{N\times M}$ is the Green's function-based channel between the $p$ polarization direction of the receiver with $N$ antennas and the $q$ polarization direction of the transmitter with $M$ antennas. Note that the $(n,m)$-th element of $\mathbf{H}^{pq}$ is the $(p,q)$-th element of $\mathbf{G}\left( \mathbf{r}_{r,n},\mathbf{r}_{t,m} \right)$, i.e., $\left[ \mathbf{H}^{pq} \right] _{nm}=G_{nm}^{pq}$.

\begin{rem}
In this paper, both the scalar and dyadic Green's function-based channel models are studied with single and triple polarization, respectively. In practice, if dipoles are oriented along one specific Euclidean axis, the polarization along this axis should be considered. Indeed, if dipoles are oriented along multiple Euclidean axes, as shown in \cite[Fig. 2]{poon2005degrees}, multiple polarization needs to be considered. Moreover, a more generalized consideration of polarization in MIMO systems was extensively studied in \cite{poon2011degree}, where both transverse electric and magnetic polarization along every possible propagation direction were analyzed. The EDoF performance analysis framework with this more generalized polarization modelling is regarded as an important future direction.
\end{rem}

\begin{figure*}[t]
{{\begin{align}\tag{3}\label{Dyadic} 
\mathbf{G}\left( \mathbf{r},\mathbf{s} \right) &=\frac{1}{4\pi}\left( \mathbf{I}_3+\frac{\nabla _{\mathbf{r}}\nabla _{\mathbf{r}}^{H}}{\kappa _{0}^{2}} \right) \frac{\exp \left( -j\kappa _0\left| \mathbf{r}-\mathbf{s} \right| \right)}{\left| \mathbf{r}-\mathbf{s} \right|}=\left( \mathbf{I}_3+\frac{\nabla _{\mathbf{r}}\nabla _{\mathbf{r}}^{H}}{\kappa _{0}^{2}} \right) G\left( \mathbf{r},\mathbf{s} \right)\\
&=\left( 1+\frac{j}{\kappa _0\left| \mathbf{r}-\mathbf{s} \right|}-\frac{1}{\kappa _{0}^{2}\left| \mathbf{r}-\mathbf{s} \right|^2} \right) \mathbf{I}_3G\left( \mathbf{r},\mathbf{s} \right) +\left( \frac{3}{\kappa _{0}^{2}\left| \mathbf{r}-\mathbf{s} \right|^2}-\frac{3j}{\kappa _0\left| \mathbf{r}-\mathbf{s} \right|}-1 \right) \vec{\mathbf{a}}_{rs}\vec{\mathbf{a}}_{rs}^TG\left( \mathbf{r},\mathbf{s} \right)\notag
\end{align}}
\hrulefill
\vspace*{-0.6cm}
%\vspace*{3pt}
}\end{figure*}

\begin{figure*}[t]
{{\begin{align}\tag{8} \label{EDoFterms_Scalar}
\begin{cases}
	\mathrm{tr}\left( \mathbf{R}_S \right) =\sum_{m=1}^M{\sum_{n=1}^N{\left[ \mathbf{R}_S \right] _{mm}^{2}}}=\sum_{m=1}^M{\sum_{n=1}^N{\left| G_{nm} \right|^2}}\\
	\left\| \mathbf{R}_S \right\| _{\mathrm{F}}^{2}=\sum_{m_1=1}^M{\sum_{m_2=1}^N{\left[ \mathbf{R}_S \right] _{m_1m_2}^{2}}}=\sum_{m_1=1}^M{\sum_{m_2=1}^N{\left| \sum_{n=1}^N{G_{nm_1}^{*}G_{nm_2}} \right|^2}}\\
\end{cases}
\end{align}}
\hrulefill
\vspace*{-0.6cm}
%\vspace*{3pt}
}\end{figure*}

\begin{figure*}[t]
{{\begin{align}\tag{9} \label{UPA_closed}
\varepsilon _S=\frac{D^4\left| \sum_{m=1}^M{\sum_{n=1}^N{\frac{1}{D^2+( -\frac{L_{r,H}}{2}+t( n ) \Delta _{r,H}+\frac{L_{t,H}}{2}-i( m ) \Delta _{t,H} ) ^2+(-\frac{L_{r,V}}{2}+\Delta _{r,V}k\left( n \right) +\frac{L_{t,V}}{2}-j\left( m \right) \Delta _{t,V})^2}}} \right|^2}{\sum_{m_1=1}^M{\sum_{m_2=1}^M{\left| \sum_{n=1}^N{e^{-j\frac{\kappa _0}{D}[ \Delta _{t,H}( i( m_1 ) -i( m_2 ) ) ( -\frac{L_{r,H}}{2}+t( n ) \Delta _{r,H} ) +( -\frac{L_{r,V}}{2}+\Delta _{r,V}k( n ) ) ( j( m_1 ) -j( m_2 ) ) \Delta _{t,V} ]}} \right|^2}}}
\end{align}}
\hrulefill
\vspace*{-0.6cm}
%\vspace*{3pt}
}\end{figure*}

\begin{figure*}[t]
{{\begin{align} \tag{18}\label{Numerator}
\gamma &= \mu_0 \frac{2L_{t,H}L_{r,H}}{L_{H,\max}}T\left( \frac{\left| L_{t,H}-L_{r,H} \right|}{2} \right) +\mu_0 \left( L_{t,H}+L_{r,H} \right) T\left( \frac{L_{t,H}+L_{r,H}}{2} \right) -\mu_0 \left( L_{t,H}+L_{r,H} \right) T\left( \frac{\left| L_{t,H}-L_{r,H} \right|}{2} \right)\\ &-2\mu_0 Q\left( \frac{L_{t,H}+L_{r,H}}{2} \right) +2\mu_0 Q\left( \frac{\left| L_{t,H}-L_{r,H} \right|}{2} \right)\notag
\end{align}}
\hrulefill
\vspace*{-0.65cm}
%\vspace*{3pt}
}\end{figure*}

\subsection{EDoF Performance Analysis}\label{UPA_EDOF_Ana}
In this subsection, we analyze the EDoF performance for the UPA-based XL-MIMO system with infinitely thin dipoles over the scalar or dyadic Green's function-based channels, which can provide the basis for the analysis framework for other XL-MIMO designs. Similar results have also been derived in \cite{[65]}, so we introduce them in a brief manner. More importantly, we compute EDoF expressions over the scalar Green's function-based channel in the closed-form.
\subsubsection{Scalar Green's function-based Channel}
Based on the channel as shown in \eqref{Discrete_Scalar_mn}, we can obtain the scalar Green's function-based channel correlation matrix $\mathbf{R}_S=\mathbf{H}_{S}^{H}\mathbf{H}_S\in \mathbb{C} ^{M\times M}$. As extensive simulations have validated \cite{ouyangEDoF,2024arXiv240105900L,2023arXiv230406141X}, the ordered eigenvalues of channel $\mathbf{H}_{S}$ exhibit a two-stage characteristic. In the first stage, the singular values remain approximately constant, and then, at one particular singular value, with its index defined as $\mathrm{EDoF}$, they sharply degrade, approaching to zero. That is, $\sigma _1\approx \sigma _2\approx \dots \approx \sigma _{\mathrm{EDoF}_1}\gg \sigma _{\mathrm{EDoF}_1+1}>\dots >\sigma _{\mathrm{DoF}}$. Exploiting this characteristic, we have $\mathrm{tr}( \mathbf{R}_S ) =\sum_{n=1}^{\mathrm{DoF}}{\sigma_{n}^2}\approx \mathrm{EDoF}\times \sigma _{\mathrm{EDoF}^2}$ and $\| \mathbf{R}_S \| _{\mathrm{F}}^{2}=\sum_{n=1}^{\mathrm{DoF}}{\sigma _{n}^{4}}\approx \mathrm{EDoF}\times \sigma _{\mathrm{EDoF}}^{4}$. Thus, in such case, the EDoF can be approximately computed as
\begin{equation}\label{EDoF_Scalar_Dis}
\mathrm{EDoF}\approx \frac{\left( \sum_{n=1}^{\mathrm{DoF}}{\sigma _{n}^2} \right)^2}{\sum_{n=1}^{\mathrm{DoF}}{\sigma _{n}^{4}}}=\frac{\mathrm{tr}^2\left( \mathbf{R}_S \right)}{\left\| \mathbf{R}_S \right\| _{\mathrm{F}}^{2}}\triangleq \varepsilon _S
\end{equation}
As introduced in Section~\ref{intro}, \eqref{EDoF_Scalar_Dis} is an approximate mathematical solution for the number of dominant singular values or sub-channels, which corresponds to the rigorous definitions in classical related works on the DoF \cite{franceschetti2017wave,poon2005degrees,franceschetti2015landau,pizzo2022landau}.

\begin{rem}\label{verdu}
The approximate computation of EDoF over the dyadic Green's function-based channel can be also easily derived following the similar method as $\varepsilon _S$ and is therefore omitted in the following. It is worth noting that this approximate EDoF expression is coincidentally similar as the expression introduced in \cite[Eq. (8)]{4418491} based on the theory in \cite{1003824}. However, the derivation in \cite{4418491} is implemented based on the low-SNR regime. Thus, more generalized derivation, implemented from the perspective of the singular value like the proof in this paper, is highly advocated and applicable.
\end{rem}

Terms in \eqref{EDoF_Scalar_Dis} can be expanded as \eqref{EDoFterms_Scalar}, where $\left[ \mathbf{R}_S \right] _{m_1m_2}=\sum_{n=1}^N{G_{nm_1}^{*}G_{nm_2}}$ is the $(m_1,m_2)$-th element of $\mathbf{R}_S$ with $m_1,m_2\in \left\{ 1,\dots ,M \right\} $. Based on \eqref{EDoFterms_Scalar}, \eqref{EDoF_Scalar_Dis} can be computed in the closed-form as in the following theorem.
\begin{thm}\label{UPAclosed}
For the UPA-based XL-MIMO system over the scalar Green's function-based channel, the EDoF in \eqref{EDoF_Scalar_Dis} can be computed in closed-form as \eqref{UPA_closed}.
\end{thm}

\begin{IEEEproof}
The proof of Theorem~\ref{UPAclosed} is given in Appendix~\ref{UPAClosedProof}.
\end{IEEEproof}

\subsubsection{Dyadic Green's function-based Channel}
Similar to that of the scalar Green's function, the EDoF for the UPA-based XL-MIMO system over the dyadic Green's function-based channel can be approximately computed as
\addtocounter{equation}{2}
\begin{equation}\label{EDoF_DyadicW_Dis}
\varepsilon _D=\frac{\mathrm{tr}^2\left( \mathbf{R}_D \right)}{\left\| \mathbf{R}_D \right\| _{\mathrm{F}}^{2}},
\end{equation}
where $\mathbf{R}_D=\mathbf{H}_{D}^{H}\mathbf{H}_D\in \mathbb{C} ^{3M\times 3M}$ is the dyadic Green's function-based channel correlation matrix.
\begin{rem}\label{dosinpolar}
The channel in \eqref{HD_Dis} embraces full triple polarization, constructed by the $x,y,z$ polarization directions of the transceiver. The double and single polarized channel can also be obtained from \eqref{HD_Dis}. Unless otherwise specified, the channel with double polarization includes $x$ and $y$ polarization of the transceiver, i.e.,
\begin{equation}\label{Double_Polar}
\mathbf{H}_{D,2}=\left[ \begin{matrix}
	\mathbf{H}^{xx}&		\mathbf{H}^{xy}\\
	\mathbf{H}^{yx}&		\mathbf{H}^{yy}\\
\end{matrix} \right]\in \mathbb{C} ^{2N\times 2M}.
\end{equation}
For the single polarized channel, only $x$ polarization direction of the transceiver is impinged, i.e., $\mathbf{H}_{D,1}=\left[ \mathbf{H}^{xx} \right] \in \mathbb{C} ^{N\times M}$. When only the single polarization is considered, the dyadic Green's function-based channel is boiled down into the scalar Green's function-based channel except with some additional coefficients. Note that the EDoF for the UPA-based XL-MIMO system over the double or single polarized channel can be derived by substituting the double polarized channel $\mathbf{H}_{D,2}$ and the single polarized channel $\mathbf{H}_{D,1}$ into \eqref{EDoF_DyadicW_Dis}.
\end{rem}

\section{2D CAP Plane-Based XL-MIMO}\label{2DCAP}
%\begin{figure}[t]
%\centering
%\includegraphics[scale=0.45]{FIG_System_CAP.eps}
%\caption{A 2D CAP plane-based XL-MIMO system.\label{00}}
%\vspace{-0.4cm}
%\end{figure}
\subsection{System Model}
In this section, we study the 2D CAP plane-based XL-MIMO system, where a single transmitter and a single receiver are considered. Note that the transmitter is located on the $x-y$ plane with its center being the origin. The receiver is parallel to the transmitter with its center located in $(0,0,D)$. For the transmitter, the horizontal and the vertical side-lengths are $L_{t,H}$ and $L_{t,V}$, respectively. The horizontal and vertical side-lengths for the receiver are $L_{r,H}$ and $L_{r,V}$, respectively. Let $S_T$ and $S_R$ denote the transmitting and receiving regions, respectively. We also study the scalar and dyadic Green's function-based channels for the 2D CAP plane-based XL-MIMO system are also studied. Note that the CAP plane has the continuous aperture. For a certain transmitting point $\mathbf{r}_t\in \mathbb{R} ^3$ in a continuous transmitting region $S_T$ and a certain receiving point $\mathbf{r}_r\in \mathbb{R} ^3$ in a continuous receiving region $S_R$, the scalar Green's function-based channel $G\left( \mathbf{r}_r,\mathbf{r}_t \right)$ and dyadic Green's function-based channel $\mathbf{G}\left( \mathbf{r}_{r},\mathbf{r}_{t} \right)\in \mathbb{C} ^{3\times 3}$ are

\begin{align}
G\left( \mathbf{r}_r,\mathbf{r}_t \right) & =\frac{1}{4\pi}\frac{\exp \left( -j\kappa _0\left| \mathbf{r}_r-\mathbf{r}_t \right| \right)}{\left| \mathbf{r}_r-\mathbf{r}_t \right|}, \label{CAPScalar}\\
\mathbf{G}\left( \mathbf{r}_r,\mathbf{r}_t \right) &=\left( \mathbf{I}_3+\frac{\nabla _{\mathbf{r}_r}\nabla _{\mathbf{r}_r}^{H}}{\kappa _{0}^{2}} \right) G\left( \mathbf{r}_r,\mathbf{r}_t \right) .\label{CAP_Dyadic}
\end{align}
\begin{rem}
Similar to the scenario with the UPA-based XL-MIMO system, the double and single polarized channels for the 2D CAP plane-based XL-MIMO system can be built over the $x,y$ polarization directions and only the $x$ polarization direction, respectively. The double and single polarized channel can be derived from the full polarized channel defined in \eqref{CAP_Dyadic}, by extracting $x,y$ polarized components and only $x$ polarized component from $\mathbf{G}\left( \mathbf{r}_{r},\mathbf{r}_{t} \right)$, respectively.
\end{rem}
\vspace{-0.4cm}

\subsection{EDoF Performance Analysis}
The EDoF performance for the 2D CAP plane-based XL-MIMO system over the scalar or dyadic Green's function-based channel is studied.
\subsubsection{Scalar Green's function-based Channel}\label{CAPSCA}
According to the methods in \cite{2023arXiv230406141X,[29],EDOFTVT}, we first define the auto-correlation kernel to describe the correlation characteristic for two arbitrary transmitting points $\mathbf{r}_t\in S_T$ and $\mathbf{r}_{t^{\prime}}\in S_T$  as
\begin{equation}\label{Kernel_Scalar}
K(\mathbf{r}_t,\mathbf{r}_{t^{\prime}})=\int_{S_R}{G^*\left( \mathbf{r}_r,\mathbf{r}_t \right) G\left( \mathbf{r}_r,\mathbf{r}_{t^{\prime}} \right)}d\mathbf{r}_r.
\end{equation}
Based on $K(\mathbf{r}_t,\mathbf{r}_{t^{\prime}})$, by applying the asymptotic analysis to \eqref{EDoF_Scalar_Dis} by letting $M\rightarrow \infty $, $N\rightarrow \infty$ with an invariant physical size, we can derive the EDoF for the 2D CAP plane-based XL-MIMO system over the scalar Green's function-based channel. More specifically, by letting $M,N\rightarrow \infty $, $\mathrm{tr}\left( \mathbf{R}_S \right) $, and $\left\| \mathbf{R}_S \right\| _{\mathrm{F}}^{2}$ can be represented as
\begin{align}\label{EDoFterms_Scalar_CAP}
\begin{cases}
	\mathrm{tr}\left( \mathbf{R}_S \right) \rightarrow \frac{N_HN_VM_HM_V}{L_{r,H}L_{r,V}L_{t,H}L_{t,V}}\int_{S_T}{\int_{S_R}{\left| G\left( \mathbf{r}_r,\mathbf{r}_t \right) \right|^2}d\mathbf{r}_r}d\mathbf{r}_t\\
	\left\| \mathbf{R}_S \right\| _{\mathrm{F}}^{2}\rightarrow \left( \frac{N_HN_VM_HM_V}{L_{r,H}L_{r,V}L_{t,H}L_{t,V}} \right) ^2\int_{S_T}{\int_{S_T}{\left| K(\mathbf{r}_t,\mathbf{r}_{t^{\prime}}) \right|^2}d}\mathbf{r}_td\mathbf{r}_{t^{\prime}}\\
\end{cases}
\end{align}
according to the asymptotic representation $d\mathbf{r}_r=dx_rdy_r\rightarrow \frac{L_{r,H}L_{r,V}}{N_HN_V}$, $d\mathbf{r}_{t}=dx_{t}dy_{t}\rightarrow \frac{L_{t,H}L_{t,V}}{M_HM_V}$, and $d\mathbf{r}_{t^{\prime}}=dx_{t^{\prime}}dy_{t^{\prime}}\rightarrow \frac{L_{t,H}L_{t,V}}{M_HM_V}$ \cite{[29]}. Based on the above asymptotic terms in \eqref{EDoFterms_Scalar_CAP}, we can derive the EDoF expressions for the CAP-based XL-MIMO system over the scalar Green's function-based channel as in the following corollary.

\begin{coro}\label{CAP_EDoF_Scalar_Coro}
The EDoF for the 2D CAP plane-based XL-MIMO system over the scalar Green's function-based channel can be calculated as
\begin{equation}\label{CAP_EDoF_Scalar}
\varPsi _S=\frac{\left( \int_{S_T}{\int_{S_R}{\left| G\left( \mathbf{r}_r,\mathbf{r}_t \right) \right|^2}d\mathbf{r}_r}d\mathbf{r}_t \right) ^2}{\int_{S_T}{\int_{S_T}{\left| K(\mathbf{r}_t,\mathbf{r}_{t^{\prime}}) \right|^2}d}\mathbf{r}_td\mathbf{r}_{t^{\prime}}}.
\end{equation}
\end{coro}

\begin{rem}
Note that $\int_{S_T}{\int_{S_R}{\left| G\left( \mathbf{r}_r,\mathbf{r}_t \right) \right|^2}d\mathbf{r}_r}d\mathbf{r}_t $ and ${\int_{S_T}{\int_{S_T}{\left| K(\mathbf{r}_t,\mathbf{r}_{t^{\prime}}) \right|^2}d}\mathbf{r}_td\mathbf{r}_{t^{\prime}}}$ denote the overall channel gain across the transmitting and receiving regions and the overall power of the correlation kernel function $K(\mathbf{r}_t,\mathbf{r}_{t^{\prime}})$ across the transmitting region.
\end{rem}

When considering the scalar Green's function-based channel, the EDoF for the 2D CAP plane-based XL-MIMO system can be computed in closed-form as in the following theorem.
\begin{thm}\label{CAPclosed}
The EDoF for the 2D CAP plane-based XL-MIMO system over the scalar Green's function-based channel can be approximately calculated in closed-form as
\begin{equation}\label{CAP_EDoF_Scalar_Closed}
\varPsi _{S,\mathrm{c}}=\frac{\gamma ^2}{\xi},
\end{equation}
where $\gamma$ is given in \eqref{Numerator} with $\mu_0 =\left( 1/4\pi \right) ^2$ and $L_{H,\max}=\max \left\{ L_{t,H},L_{r,H} \right\}$. Note that $T\left( x \right)$ is
\addtocounter{equation}{1}
\begin{equation}\label{T_function}
\begin{aligned}
T\left( x \right)& =\frac{2L_{t,V}L_{r,V}}{D}\arctan{\frac{x}{D}}+x\ln \left( \frac{\mu _1+4x^2}{\mu _2+4x^2} \right)
\\
&+\sqrt{\mu _1}\arctan{\frac{2x}{\sqrt{\mu _1}}}-\sqrt{\mu _2}\arctan{\frac{2x}{\sqrt{\mu _2}}},
\end{aligned}
\end{equation}
where $\mu _1=\left( L_{t,V}-L_{r,V} \right) ^2+4D^2$ and $\mu _2=\left( L_{t,V}+L_{r,V} \right) ^2+4D^2$. As for $Q\left( x \right)$, we have
\begin{equation}\label{Q_function}
\begin{aligned}
Q\left( x \right) &=L_{t,V}L_{r,V}\ln \left( D^2+x^2 \right) +\frac{4x^2+\mu _1}{8}\ln \left( \mu _1+4x^2 \right)
\\
&-\frac{4x^2+\mu _2}{8}\ln \left( \mu _2+4x^2 \right).
\end{aligned}
\end{equation}
%As for $\xi$, we have
%\begin{equation}\label{denominator_close}
%\xi =\frac{L_{t,H}L_{r,H}L_{t,V}L_{r,V}}{\left( M_sN_s \right) ^2}\sum_{k=1}^{M_s}{\sum_{j=1}^{M_s}{\left( \sum_{i=1}^{N_s}{\zeta _{kji}} \right) ^2}},
%\end{equation}
%where $\zeta _{kji}=G^*\left( \mathbf{r}_{r,i},\mathbf{r}_{t,k} \right) G\left( \mathbf{r}_{r,i},\mathbf{r}_{t,j} \right) $ with $M_s$ and $N_s$ being the sampling numbers in the transmitting region and the receiving region, respectively. Besides, $\mathbf{r}_{r,i}$, $\mathbf{r}_{t,k}$, and $\mathbf{r}_{t,j}$ denote the coordinate positions for the $i$-th sampled receiving point, the $k$-th sampled transmitting point, and the $j$-th sampled transmitting antenna, respectively.
As for $\xi$, we have
\begin{equation}\label{denominator_close}
\begin{aligned}
\xi &=\frac{4\mu _3\varphi L_{t,H}^{2}L_{t,V}^{2}}{D^2\left( 4D^2+L_{t,H}^{2} \right)}+\frac{2\mu _3\varphi L_{t,H}L_{t,V}^{2}}{D^3}\arctan{\frac{L_{t,H}}{2D}}\\
&+\frac{16\mu _3\varphi L_{t,V}^{2}}{4D^2+L_{t,H}^{2}}-\frac{4\mu _3\varphi L_{t,V}^{2}}{D^2}
\end{aligned}
\end{equation}
where $\mu _3=\left( \mu _0L_{r,V}L_{r,H} \right) ^2$ and  $\varphi$ is the phase-related coefficient, which is computed as
\begin{equation}\label{expp}
\varphi \!=\!\frac{1}{N_{s}^{2}M_{s}^{2}}\sum_{o=1}^{M_s}{\sum_{u=1}^{M_s}{| \!\!\sum_{k^{\prime}=1}^{N_s}\!{\exp \left( j\kappa _0d_{k^{\prime}o} \right) \exp \left( -j\kappa _0d_{k^{\prime}u}^{\prime} \right)} |^2}},
\end{equation}
where $d_{k^{\prime}o}=\left| \mathbf{r}_{r,k^{\prime}}-\mathbf{r}_{t,o} \right|$ and $d_{k^{\prime}u}^{\prime}=\left| \mathbf{r}_{r,k^{\prime}}-\mathbf{r}_{t^{\prime},u} \right|$ are the distances between the $k^{\prime}$-th uniformly sampled receiving point and the $o$-th or $u$-th uniformly sampled transmitting points, respectively, with $\mathbf{r}_{r,k^{\prime}}\in S_R$, $\left\{ \mathbf{r}_{t,o}, \mathbf{r}_{t^{\prime},u} \right\} \in S_T$, $\left\{ o,u \right\} =\left\{ 1,\dots ,M_s \right\}$, $k^{\prime}=\left\{ 1,\dots ,N_s \right\} $. $M_s$ is the number of uniformly sampled transmitting points, and $N_s$ is the number of uniformly sampled receiving points.
\vspace{-0.25cm}

\end{thm}

\begin{IEEEproof}
The detailed proof is provided in Appendix~\ref{gamma}.
\end{IEEEproof}

\begin{rem}\label{remark4}
As observed in Theorem~\ref{CAPclosed}, the EDoF performance for 2D CAP plane-based XL-MIMO systems is determined by the physical size of the transceiver, transmitting distance, and carrier frequency (or the wavenumber). Meanwhile, as shown in  \eqref{Numerator}, the overall channel gain $\int_{S_T}{\int_{S_R}{\left| G\left( \mathbf{r}_r,\mathbf{r}_t \right) \right|^2}d\mathbf{r}_r}d\mathbf{r}_t $ depends only on the physical size and transmitting distance. Besides, when the physical size of the transmitting plane is much larger than that of the receiving plane, Theorem~\ref{CAPclosed} can be further simplified. When $L_{t,\{ V,H \}}\gg L_{r,\{ V,H \}}$, $\gamma$ in \eqref{Numerator} can be approximated as $\gamma \approx 2L_{r,H}\mu _0T( \frac{L_{t,H}}{2} ) $ by applying the fact that $\frac{| L_{t,H}-L_{r,H} |}{2}\approx \frac{L_{t,H}}{2}$, $\frac{L_{t,H}+L_{r,H}}{2}\approx \frac{L_{t,H}}{2}$, and $L_{H,\max}=L_{t,H}$. Thus, closed-form results in \eqref{CAP_EDoF_Scalar_Closed} can be further formulated as
\begin{equation}
\varPsi _{S,\mathrm{c}}\approx \frac{T^2( \frac{L_{t,H}}{2} )}{\left[ \begin{array}{c}
	\frac{\varphi L_{t,H}^{2}L_{t,V}^{2}L_{r,V}^{2}}{D^2\left( 4D^2+L_{t,H}^{2} \right)}+\frac{\varphi L_{t,H}L_{t,V}^{2}L_{r,V}^{2}}{2D^3}\arctan{\frac{L_{t,H}}{2D}}\\
	+\frac{4\varphi L_{t,V}^{2}L_{r,V}^{2}}{4D^2+L_{t,H}^{2}}-\frac{\varphi L_{t,V}^{2}L_{r,V}^{2}}{D^2}\\
\end{array} \right]}.
\end{equation}
\end{rem}

%\vspace{-0.3cm}

\begin{figure*}[t]
{{\begin{align}\tag{27} \label{Patch_region}
V_{T,m}=\{ ( x_{t}^{m},y_{t}^{m},0 ) \in \mathbb{R} ^3:\{ \mathbf{r}_{t,m}\left( 1 \right) -\frac{A_{t,H}}{2}\leqslant x_{t}^{m}\leqslant \mathbf{r}_{t,m}\left( 1 \right) +\frac{A_{t,H}}{2};\mathbf{r}_{t,m}\left( 2 \right) -\frac{A_{t,V}}{2}\leqslant y_{t}^{m}\leqslant \mathbf{r}_{t,m}\left( 2 \right) +\frac{A_{t,V}}{2} \} \}
\\
V_{R,n}=\{ ( x_{r}^{n},y_{r}^{n},D ) \in \mathbb{R} ^3:\{ \mathbf{r}_{r,n}\left( 1 \right) -\frac{A_{r,H}}{2}\leqslant x_{r}^{n}\leqslant \mathbf{r}_{r,n}\left( 1 \right) +\frac{A_{r,H}}{2};\mathbf{r}_{r,n}\left( 2 \right) -\frac{A_{r,V}}{2}\leqslant y_{r}^{n}\leqslant \mathbf{r}_{r,n}\left( 2 \right) +\frac{A_{r,V}}{2} \} \}\notag
\end{align}}
\hrulefill
\vspace*{-0.6cm}
%\vspace*{3pt}
}\end{figure*}

\begin{figure*}[t]
{{\begin{align}\tag{29} \label{Patch_scalar}
\psi _S=\frac{\left\{ \sum_{n=1}^N{\sum_{m=1}^M{\left[ \int_{V_{T,m}}{\int_{V_{R,n}}{\left| G\left( \mathbf{r}_{r}^{n},\mathbf{r}_{t}^{m} \right) \right|^2}}d\mathbf{r}_{r}^{n}d\mathbf{r}_{t}^{m} \right]}} \right\} ^2}{\sum_{m_1=1}^M{\sum_{m_2=1}^M{\int_{V_{T,m_1}}{\int_{V_{T,m_2}}{\left| \sum_{n=1}^N{\left[ \int_{V_{R,n}}{G^*\left( \mathbf{r}_{r}^{n},\mathbf{r}_{t}^{m_1} \right) G\left( \mathbf{r}_{r}^{n},\mathbf{r}_{t^{\prime}}^{m_2} \right)}d\mathbf{r}_{r}^{n} \right]} \right|^2}}d\mathbf{r}_{t}^{m_1}d\mathbf{r}_{t^{\prime}}^{m_2}}}}
\end{align}}
\hrulefill
\vspace*{-0.6cm}
%\vspace*{3pt}
}\end{figure*}

\begin{figure*}[t]
{{\begin{align}\tag{31} \label{Patch_Dyadic}
\psi _D=\frac{\left( \sum\nolimits_{n=1}^N{\sum\nolimits_{m=1}^M{\sum\nolimits_{l=1}^3{\sum\nolimits_{p=1}^3{\left[ \int_{V_{T,m}}{\int_{V_{R,n}}{\left| G^{lp}\left( \mathbf{r}_{r}^{n},\mathbf{r}_{t}^{m} \right) \right|^2}}d\mathbf{r}_{r}^{n}d\mathbf{r}_{t}^{m} \right]}}}} \right) ^2}{\sum\nolimits_{m_1=1}^M{\sum\nolimits_{m_2=1}^M{\sum\nolimits_{p=1}^3{\sum\nolimits_{q=1}^3{\int_{V_{T,m_1}}{\int_{V_{T,m_2}}{\left| \sum\nolimits_{n=1}^N{\sum\nolimits_{l=1}^3{\left[ \int_{V_{R,n}}{\!}\!\left[ G^{lp}\left( \mathbf{r}_{r}^{n},\mathbf{r}_{t}^{m_1} \right) \right] ^*G^{lq}\left( \mathbf{r}_{r}^{n},\mathbf{r}_{t^{\prime}}^{m_2} \right) d\mathbf{r}_{r}^{n} \right]}} \right|^2d\mathbf{r}_{t}^{m_1}d\mathbf{r}_{t^{\prime}}^{m_2}}}}}}}}.
\end{align}}
\hrulefill
\vspace*{-0.6cm}
%\vspace*{3pt}
}\end{figure*}

\subsubsection{Dyadic Green's function-based Channel}
To compute the EDoF performance for the 2D CAP plane-based XL-MIMO system over the dyadic Green's function-based channel, we define the polarized auto-correlation kernel function over a specific polarization direction pair $\left\{ p,q \right\} \in \left\{ 1,2,3 \right\} $ as
\begin{equation}\label{Kernel_Dyadic}
\hat{K}(\mathbf{r}_t,\mathbf{r}_{t^{\prime}},p,q)\!=\!\sum_{l=1}^3{\int_{S_R}{\!}\!\left[ G^{lp}\left( \mathbf{r}_r,\mathbf{r}_t \right) \right] ^*G^{lq}\left( \mathbf{r}_r,\mathbf{r}_{t^{\prime}} \right) d\mathbf{r}_r},
\end{equation}
where the polarization subscripts are represented, for simplicity, as $x\rightarrow 1$, $y\rightarrow 2$, and $z\rightarrow 3$, respectively. Besides, $G^{lp}\left( \mathbf{r}_r,\mathbf{r}_t \right)$ denotes the $(l,p)$-th element of $\mathbf{G}\left( \mathbf{r}_{r},\mathbf{r}_{t} \right)$ as given in \eqref{CAP_Dyadic}. Based on the asymptotic analysis by letting $M\rightarrow \infty $, $N\rightarrow \infty$ to \eqref{EDoF_DyadicW_Dis} with an invariant physical size, we derive the EDoF for the 2D CAP plane-based XL-MIMO over the dyadic Green's function-based channel as in the following corollary.

\begin{coro}\label{CAP_EDoF_Dyadic_Coro}
The EDoF for the 2D CAP plane-based XL-MIMO system over the dyadic Green's function-based channel can be calculated as
\begin{equation}\label{CAP_EDoF_Dyadic}
\varPsi _D=\frac{\left( \sum_{l=1}^3{\sum_{p=1}^3{\int_{S_T}{\int_{S_R}{\left| G^{lp}\left( \mathbf{r}_r,\mathbf{r}_t \right) \right|^2d\mathbf{r}_r}d\mathbf{r}_t}}} \right) ^2}{\sum_{p=1}^3{\sum_{q=1}^3{\int_{S_T}{\int_{S_T}{\left| \hat{K}(\mathbf{r}_t,\mathbf{r}_{t^{\prime}},p,q) \right|}^2d}\mathbf{r}_td\mathbf{r}_{t^{\prime}}}}}.
\end{equation}
\end{coro}
\begin{IEEEproof}
The detailed proof can be found in \cite{EDOFTVT} and is therefore omitted.
\end{IEEEproof}

\begin{rem}
Note that the EDoF computation in Corollary~\ref{CAP_EDoF_Dyadic_Coro} includes full triple polarization. However, for the double or single polarized channel as discussed in Remark~\ref{dosinpolar}, the EDoF performance can also be studied based on the analysis framework in Corollary~\ref{CAP_EDoF_Dyadic_Coro}. It is worth noticing that the EDoF performance analysis framework in Corollary~\ref{CAP_EDoF_Dyadic_Coro} holds for the channel with arbitrary numbers of polarization. We can compute the EDoF for the 2D CAP plane-based over the channel with arbitrary numbers of polarization as
\begin{equation}\label{CAP_EDoF_Dyadic_arbi}
\varPsi _D=\frac{\left( \sum_{l=1}^{N_p}{\sum_{p=1}^{N_p}{\int_{S_T}{\int_{S_R}{\left| G^{lp}\left( \mathbf{r}_r,\mathbf{r}_t \right) \right|^2d\mathbf{r}_r}d\mathbf{r}_t}}} \right) ^2}{\sum_{p=1}^{N_p}{\sum_{q=1}^{N_p}{\int_{S_T}{\int_{S_T}{\left| \hat{K}(\mathbf{r}_t,\mathbf{r}_{t^{\prime}},p,q) \right|^2}d}\mathbf{r}_td\mathbf{r}_{t^{\prime}}}}},
\end{equation}
where $\hat{K}(\mathbf{r}_t,\mathbf{r}_{t^{\prime}},p,q)\!=\!\sum_{l=1}^{N_p}{\int_{S_R}{\!}\!\left[ G^{lp}\left( \mathbf{r}_r,\mathbf{r}_t \right) \right] ^*G^{lq}\left( \mathbf{r}_r,\mathbf{r}_{t^{\prime}} \right) d\mathbf{r}_r}$ and $N_p=\left\{ 1,2,3 \right\}$  is the number of channel polarization. When $N_p=1$, \eqref{CAP_EDoF_Dyadic_arbi} converts to \eqref{CAP_EDoF_Scalar} over the scalar Green's function-based channel.
\end{rem}

\section{Other XL-MIMO Hardware Designs}\label{other}
In this section, we introduce the EDoF performance analysis framework for other XL-MIMO hardware designs, UPA-based XL-MIMO with patch antennas and 1D XL-MIMO designs.
%\begin{figure}[t]
%\centering
%\includegraphics[scale=0.2]{FIG_System_UPA_Patch.eps}
%\caption{An illustration of the transmitter equipped with UPA-based XL-MIMO with patch antennas. \label{patch}}
%\vspace{-0.4cm}
%\end{figure}

\subsection{UPA-Based XL-MIMO with Patch Antennas}
The UPA-based XL-MIMO investigated in Section~\ref{point} views each antenna element as an infinitely thin dipole. Another practical UPA-based XL-MIMO design is to regard each antenna element as a patch antenna with a certain size and continuous aperture in each antenna element \cite{ZheSurvey,[97]}. For this UPA-based XL-MIMO design with patch antennas, across each antenna element, the EM wave should be captured with the aid of exact integrals due to the continuous characteristic for each element. We assume that the physical sizes of the transmitting and receiving patch antenna element can be denoted as $A_{t,H}\times A_{t,V}$ and $A_{r,H}\times A_{r,V}$, respectively, where $A_{t,H}$, $A_{t,V}$, $A_{r,H}$, and $A_{r,V}$ denote the horizontal or vertical element side-length of the transmitting antenna element and the horizontal or vertical element side length of the receiving antenna element, respectively. Other parameter definitions are similar to those of the UPA-based XL-MIMO system in Section~\ref{point}. Moreover, we denote the transmitting region and the receiving region as $V_T=\left\{ V_{T,m}:m=1,\dots ,M \right\} $ and $V_R=\left\{ V_{R,n}:n=1,\dots ,N \right\} $, where $V_{T,m}$ and $V_{R,n}$ are the regions of the $m$-th transmitting patch antenna and the $n$-th receiving patch antenna, respectively, defined as shown in \eqref{Patch_region} with $\mathbf{r}_{t,m}$ and $\mathbf{r}_{r,n}$ being the coordinates of centers of the $m$-th transmitting and the $n$-th receiving patch antennas, respectively. Note that $\mathbf{r}_{t,m}$ and $\mathbf{r}_{r,n}$ are also the coordinates of the $m$-th transmitting and the $n$-th receiving infinitely thin dipoles, respectively, as defined in Section~\ref{point}.

When computing the EDoF for the UPA-based XL-MIMO system with patch antennas, as discussed above, exact integrals across each patch antenna element should be computed. The auto-correlation kernel function $K(\mathbf{r}_t,\mathbf{r}_{t^{\prime}})$ can be derived from \eqref{Kernel_Scalar}, by considering ${V_T}$ and $V_R$ in \eqref{Patch_region}, as
\addtocounter{equation}{1}
$K(\mathbf{r}_t,\mathbf{r}_{t^{\prime}})=\int_{V_R}{G^*\left( \mathbf{r}_r,\mathbf{r}_t \right) G\left( \mathbf{r}_r,\mathbf{r}_{t^{\prime}} \right)}d\mathbf{r}_r=\sum_{n=1}^N{[ \int_{V_{R,n}}{G^*( \mathbf{r}_{r}^{n},\mathbf{r}_t ) G( \mathbf{r}_{r}^{n},\mathbf{r}_{t^{\prime}} )}d\mathbf{r}_{r}^{n} ]},
$
where $\mathbf{r}_{r}^{n}=( x_{r}^{n},y_{r}^{n},D )\in V_{R,n}$ and $G( \mathbf{r}_{r}^{n},\mathbf{r}_t )$ can be derived as in \eqref{CAPScalar}. Based on the above definitions and motivated by Corollary~\ref{CAP_EDoF_Scalar_Coro}, we derive the EDoF analysis framework for the UPA-based XL-MIMO system with patch antennas over the scalar Green's function-based channel in the following.
\begin{coro}\label{Patch_EDoF_Scalar_Coro}
For the UPA-based XL-MIMO system with patch antennas over the scalar Green's function-based channel, the EDoF $\psi _S$ can be computed as
\begin{equation}\label{Patch_EDoF_Scalar}
\begin{aligned}
\psi _S=\frac{\left( \int_{V_T}{\int_{V_R}{\left| G\left( \mathbf{r}_r,\mathbf{r}_t \right) \right|^2}d\mathbf{r}_r}d\mathbf{r}_t \right) ^2}{\int_{V_T}{\int_{V_T}{\left| K(\mathbf{r}_t,\mathbf{r}_{t^{\prime}}) \right|^2}d}\mathbf{r}_td\mathbf{r}_{t^{\prime}}}.
\end{aligned}
\end{equation}
Note that \eqref{Patch_EDoF_Scalar} can be further written as \eqref{Patch_scalar}.
\end{coro}
\begin{IEEEproof}
This corollary can be easily derived by substituting $V_T$ and $V_R$ into \eqref{CAP_EDoF_Scalar}.
\end{IEEEproof}

Similarly, based on \eqref{Kernel_Dyadic}, the polarized auto-correlation kernel function for the UPA-based XL-MIMO system with patch antennas can be defined as
\addtocounter{equation}{1}
$\hat{K}(\mathbf{r}_t,\mathbf{r}_{t^{\prime}},p,q)\!=\!\sum_{l=1}^3{\int_{V_R}{\!}\![ G^{lp}( \mathbf{r}_r,\mathbf{r}_t ) ] ^*G^{lq}( \mathbf{r}_r,\mathbf{r}_{t^{\prime}} ) d\mathbf{r}_r}=\sum_{n=1}^N{\sum_{l=1}^3{[ \int_{V_{R,n}}{\!}\![ G^{lp}( \mathbf{r}_{r}^{n},\mathbf{r}_t ) ] ^*G^{lq}( \mathbf{r}_{r}^{n},\mathbf{r}_{t^{\prime}} ) d\mathbf{r}_{r}^{n} ]}}.
$
Thus, we can derive the EDoF performance over the dyadic Green's function-based channel as in the following corollary.
\begin{coro}\label{Patch_EDoF_Dyadic_Coro}
For the UPA-based XL-MIMO system with patch antennas over the dyadic Green's function-based channel, the EDoF $\psi _D$ can be computed as
\begin{equation}\label{Patch_EDoF_Dyadic}
\begin{aligned}
\psi _D=\frac{\left( \sum_{l=1}^3{\sum_{p=1}^3{\int_{V_T}{\int_{V_R}{\left| G^{lp}\left( \mathbf{r}_r,\mathbf{r}_t \right) \right|^2d\mathbf{r}_r}d\mathbf{r}_t}}} \right) ^2}{\sum_{p=1}^3{\sum_{q=1}^3{\int_{V_T}{\int_{V_T}{\left| \hat{K}(\mathbf{r}_t,\mathbf{r}_{t^{\prime}},p,q) \right|^2}d}\mathbf{r}_td\mathbf{r}_{t^{\prime}}}}}.
\end{aligned}
\end{equation}
Besides, \eqref{Patch_EDoF_Dyadic} can be further expanded as in \eqref{Patch_Dyadic}.
\end{coro}
\begin{IEEEproof}
We can easily prove this corollary by substituting $V_T$ and $V_R$ into \eqref{CAP_EDoF_Dyadic}.
\end{IEEEproof}

\begin{rem}
From a mathematical perspective, Corollary~\ref{Patch_EDoF_Scalar_Coro} and Corollary~\ref{Patch_EDoF_Dyadic_Coro} provide an important unified analytical framework due to the flexible configuration of the patch antenna element. When the side-lengths of each patch antenna element be infinitesimal, that is $A_{\left\{ t,r \right\} ,\left\{ V,H \right\}}\rightarrow 0$, or assuming that only the center of each patch antenna element is impinged for transmitting/receiving EM wave, that is $V_{T,m}=\mathbf{r}_{t,m}$ and $V_{R,n}=\mathbf{r}_{r,n}$, Corollary~\ref{Patch_EDoF_Scalar_Coro} and Corollary~\ref{Patch_EDoF_Dyadic_Coro} are equivalent to \eqref{EDoF_Scalar_Dis} and \eqref{EDoF_DyadicW_Dis} for UPA-based XL-MIMO systems with infinitely thin dipoles, respectively. Moreover, when the side-lengths of each patch antenna element equal to the antenna spacing in the respective directions, that is 
$A_{t,\left\{ H,V \right\}}=\Delta _{t,\left\{ H,V \right\}}$ and $A_{r,\left\{ H,V \right\}}=\Delta _{r,\left\{ H,V \right\}}$, the array aperture for the UPA-based XL-MIMO with patch antennas would be continuous, and thus,  Corollary~\ref{Patch_EDoF_Scalar_Coro} and Corollary~\ref{Patch_EDoF_Dyadic_Coro} would covert to Corollary~\ref{CAP_EDoF_Scalar_Coro} and Corollary~\ref{CAP_EDoF_Dyadic_Coro} for 2D CAP plane-based XL-MIMO, respectively.
\end{rem}
\subsection{1D XL-MIMO Hardware Designs}
The above investigated XL-MIMO hardware designs are all of the 2D characteristics. In this part, we investigate 1D XL-MIMO hardware designs, ULA-based XL-MIMO, and 1D CAP line segment-based XL-MIMO, and derive their respective EDoF performance analysis framework.
\subsubsection{ULA-Based XL-MIMO}
Under this hardware design, we assume that the transmitter equipped with ULA-based XL-MIMO is located on the $y$-axis with its centering coordinate being $(0,0,0)$. The side-length and the number of antennas for the transmitter can be denoted as $L_t$ and $M$, respectively. Besides, the receiver equipped with the ULA-based XL-MIMO embraces the side-length of $L_r$ and $N$ antennas. Note that the coordinates of the $m$-th transmitting antenna or $n$-th receiving antenna can be denoted as $
\mathbf{r}_{t,m}=[0,-\frac{L_{t,V}}{2}+\left( m-1 \right) \Delta _{t},0]^T\in \mathbb{R} ^3$ and $\mathbf{r}_{r,n}=[0,-\frac{L_{r,V}}{2}+\left( n-1 \right) \Delta _{r},D]^T\in \mathbb{R} ^3$, respectively, where $\Delta _t=L_t/M$ and $\Delta _r=L_r/N$ are the antenna spacing for the transmitter and the receiver, respectively, with $m=1,\dots ,M$ and $n=1,\dots ,N$.

Similar to the UPA-based XL-MIMO system with infinitely thin dipoles, we can also compute the EDoF performance for the ULA-based XL-MIMO system over the scalar and dyadic Green's function-based channel same as in \eqref{EDoF_Scalar_Dis} and \eqref{EDoF_DyadicW_Dis}, respectively, where applied channel matrices are derived by substituting the parameters of the ULA-based XL-MIMO system into
\eqref{Discrete_Scalar_mn} and \eqref{Discrete_Dyadic_mn}, respectively. Note that, when considering the scalar Green's function-based channel, we can compute the EDoF in closed-form as in the following theorem.
\begin{thm}\label{TheoremULAclosed}
The EDoF over the scalar Green's function-based channel can be computed in the closed-form as
\addtocounter{equation}{1}
\begin{equation}\label{ULAclosed}
\varepsilon _S\!=\!\frac{D^4| \sum_{m=1}^M{\sum_{n=1}^N{\frac{1}{D^2+[ -\frac{L_r}{2}+\left( n-1 \right) \Delta _r+\frac{L_t}{2}-( m-1 ) \Delta _t ] ^2}}} |^2}{\sum_{m_1=1}^M\!{\sum_{m_2=1}^M{| \sum_{n=1}^N\!{e^{-j\frac{\kappa _0}{D}[ ( -\frac{L_r}{2}+( n-1 ) \Delta _r ) ( m_1-m_2 ) \Delta _t ]}} |^2}}}.
\end{equation}
\end{thm}
\begin{IEEEproof}
The proof can be easily derived based on the similar method to that of Theorem~\ref{UPAclosed}.
\end{IEEEproof}

\subsubsection{1D CAP Line Segment-Based XL-MIMO}
For this design, unlike 2D CAP planes considered in Section~\ref{2DCAP}, we study 1D line segments with the continuous array aperture. The transmitter with the side-length of $L_t$ is located on the $y$ axis with its center being $(0,0,0)$. And the receiver with the side-length of $L_r$ is parallel to the transmitter with its center being $(0,0,D)$. Note that the EDoF performance for this design over the scalar and dyadic Green's function-based channel can be formulated similarly to that of the 2D CAP plane-based XL-MIMO system as shown in Corollary~\ref{CAP_EDoF_Scalar_Coro} and Corollary~\ref{CAP_EDoF_Dyadic_Coro}, respectively, by substituting the transmitting and receiving regions of this design into these corollaries. More importantly, over the scalar Green's function-based channel, the EDoF can also be computed in closed-form as in the following theorem.
\begin{thm}\label{Theorem1DCAPclosed}
We can compute the EDoF over the scalar Green's function-based channel in closed-form as
\begin{equation}\label{1DCAPclosed}
\varPsi _{S,\mathrm{c}}=\frac{\left\{ 2L_tL_r-D^2\ln \left[ \frac{\left( L_t+L_r \right) ^2+4D^2}{\left( L_t-L_r \right) ^2+4D^2} \right] \right\} ^2}{\varphi \left( L_tL_r \right) ^2},
\end{equation}
where $\varphi$ is similar to the definition in \eqref{expp}.
\end{thm}
\begin{IEEEproof}
We can easily obtain this theorem based on a similar method to that of Theorem~\ref{CAPclosed}.
\end{IEEEproof}
\begin{rem}
The capacity performance $C$ for all investigated XL-MIMO designs can be directly derived based on the studied EDoF expressions in this paper as $C=\mathrm{EDoF}\cdot \log _2( 1+\frac{\alpha P}{\mathrm{EDoF}^2N_0}) $ \cite{4418491}, where $\alpha$ is the overall channel gain, $P$ is the transmitting power, and $N_0$ is the noise power. 
Note that $\alpha$ can denoted as the square root of the numerator of the respective EDoF expression.
%$\mathrm{tr}( \mathbf{R} )$ and $\int{\int{\left| G\left( \mathbf{r}_r,\mathbf{r}_t \right) \right|^2}d\mathbf{r}_r}d\mathbf{r}_t $ for the discrete and continuous XL-MIMO designs, respectively.
\end{rem}
\vspace{-0.5cm}
\begin{rem}\label{farnear}
Note that all derived results in this paper hold for near-field spherical wave channel models, which are described by dyadic and scalar Green's functions with triple and single polarization directions, respectively. In the far-field scenario, a planar wave channel model is assumed, where all links embrace the same signal amplitude and the angle of arrival/departure \cite{ZheSurvey}. For the scalar Green's function-based scenario, in the far-field, $\mathbf{H}_S$ in Section~\ref{SystemUPA} has a rank of $1$ and thus the EDoF for the far-field scalar Green's function-based scenario is $1$. For the dyadic Green's function-based scenario, under the far-field approximation $\| \mathbf{r}-\mathbf{s} \| \gg \lambda$, the operator $( \mathbf{I}_3+\frac{\nabla _{\mathbf{r}}\nabla _{\mathbf{r}}^{H}}{\kappa _{0}^{2}} )$ in \eqref{Dyadic} can be approximated as $(\mathbf{I}_3-\vec{\mathbf{a}}_{rs}\vec{\mathbf{a}}_{rs}^{T})$, which has a rank of $2$ \cite{poon2005degrees}. Thus, the EDoF for the far-field dyadic Green's function-based scenario is $2$. Notably, in the near-field, spherical waves manifest different phase and signal amplitude for each link. This phenomenon can significantly increase the ranks of the channels and improve the EDoFs, which would be substantially higher than those in the far-field scenario. This argument will also be discussed in Numerical Results section.  
\end{rem}

\vspace{-0.25cm}

\section{EDoF Analysis with Mutual Coupling}\label{mutual}
With the continuing increase in the number of antennas while maintaining an invariant physical size, the antenna spacing significantly decreases and the mutual coupling effect becomes severe. In this section, we discuss the effects of the mutual coupling property on the proposed EDoF performance analysis framework. We define the mutual coupling matrices at the transmitter side with $M$ antennas and the receiver side with $N$ antennas as $\mathbf{Z}_t\in \mathbb{C} ^{M\times M}$ and $\mathbf{Z}_r\in \mathbb{C} ^{N\times N}$, respectively. Taking the modelling of $\mathbf{Z}_r$ for instance, we have \cite{balanis2016antenna}
\begin{equation}\label{MC_Matrix}
\mathbf{Z}_r=\left( Z_A+Z_L \right) \left( \mathbf{Z}_{r,C}+Z_L\mathbf{I}_N \right) ^{-1}\in \mathbb{C} ^{N\times N},
\end{equation}
where $ Z_A$ is the antenna impedance, $ Z_L$ is the load impedance, and $\mathbf{Z}_{r,C}\in \mathbb{C} ^{N\times N}$ is the  mutual impedance matrix at the receiver side, respectively. For $ Z_A$, we denote it as $Z_A=R_{Z_A}+jX_{Z_A}$, where $R_{Z_A}$ and $X_{Z_A}$ denote the resistance and the reactance, respectively. $R_{Z_A}$ and $X_{Z_A}$ can be computed as $
R_{Z_A}=\frac{\eta}{2\pi \sin ^2( \frac{\kappa d_l}{2} )}\{ \gamma _0+\ln ( \kappa d_l ) -C_i( \kappa d_l ) +\frac{\sin ( \kappa d_l )}{2}[ S_i( 2\kappa d_l ) -2S_i( \kappa d_l ) ] +\frac{\cos ( \kappa d_l )}{2}[ \gamma _0+\ln ( \frac{\kappa d_l}{2} ) +C_i( 2\kappa d_l ) -2C_i( \kappa d_l ) ] \} $ and 
$X_{Z_A}=\frac{\eta}{4\pi \sin ^2( \frac{\kappa d_l}{2} )}\{2S_i( \kappa d_l ) +\cos ( \kappa d_l ) [ 2S_i( \kappa d_l ) -S_i( 2\kappa d_l ) ]-\sin ( \kappa d_l ) [ 2C_i( \kappa d_l ) -C_i( 2\kappa d_l ) -C_i( \frac{2\kappa a^2}{d_l} ) ] \label{XZA} \}$, where $\eta $ is the intrinsic impedance, $d_l$ is the dipole length of each antenna, $a$ is the radius of the wire, $\gamma _0$ is the Euler constant, $C_i(\cdot)$ and $S_i(\cdot)$ are the cosine integral and sine integral functions, respectively.

For the UPA-based system, as illustrated in Section~\ref{point}, we can construct $\mathbf{Z}_{r,C}$ by $N_V\times N_V$ sub-matrices, that is $\mathbf{Z}_{r,C}=[ \mathbf{Z}_{r,C,pq} ] _{N_V\times N_V}$, where $\mathbf{Z}_{r,C,pq}\in \mathbb{C} ^{N_H\times N_H} $ is the impedance matrix between the $p$-th row of antennas and the $q$-th row of antennas with $\{ p,q \} =\{ 1,\dots ,N_V \}$. Moreover, we define $z_{r,n_1n_{1}^{\prime}}^{pq}$ as the $(n_1,n_{1}^{\prime})$-th element of $\mathbf{Z}_{r,C,pq}$, which represents the mutual impedance between the $n_1$-th antenna of the $p$-th row and the $n_{1}^{\prime}$-th antenna of the $q$-th row, counting from the left to right, with $\{ n_1,n_{1}^{\prime} \} =\{ 1,\dots ,N_H \}$. We define the distance, vertical distance, and horizontal distance between the $n_1$-th antenna of the $p$-th row and the $n_{1}^{\prime}$-th antenna of the $q$-th row as $d_{n_1n_{1}^{\prime}}^{pq}=\sqrt{( n_1-n_{1}^{\prime} ) ^2\varDelta _{r,H}^{2}+( p-q ) ^2\varDelta _{r,V}^{2}}$, $d_{pq,v}=| p-q | \varDelta _{r,V}$, and $d_{n_1n_{1}^{\prime},h}=| n_1-n_{1}^{\prime} | \varDelta _{r,H}$, respectively. Note that here are four possible combinations for $p,q$, and $n_1,n_{1}^{\prime}$, where the modelling of the mutual impedance $z_{r,n_1n_{1}^{\prime}}^{pq}$ for each combination is different and $z_{r,n_1n_{1}^{\prime}}^{pq}$ can be uniformly denoted as $z_{r,n_1n_{1}^{\prime}}^{pq}=R_{r,n_1n_{1}^{\prime}}^{pq}+jX_{r,n_1n_{1}^{\prime}}^{pq}$. For $p=q, n_1=n_{1}^{\prime}$, we have $z_{r,n_1n_{1}^{\prime}}^{pq}=Z_A$. For $p=q, n_1\ne n_{1}^{\prime}$, this scenario is the  ``side-by-side" scenario in \cite[Sec. 8.6.2]{balanis2016antenna}. According to \cite[Eq. 8-71]{balanis2016antenna}, we can compute $R_{r,n_1n_{1}^{\prime}}^{pq}$ and $X_{r,n_1n_{1}^{\prime}}^{pq}$ by letting the parameters in \cite[Eq. 8-71]{balanis2016antenna} be $d=d_{n_1n_{1}^{\prime},h}$ and $l=d_l$. For $p\ne q, n_1=n_{1}^{\prime}$, this scenario is the ``parallel-in-echelon" scenario in \cite[Sec. 8.6.2]{balanis2016antenna}. Based on \cite[Eq. 8-73]{balanis2016antenna}, we can compute $R_{r,n_1n_{1}^{\prime}}^{pq}$ and $X_{r,n_1n_{1}^{\prime}}^{pq}$ by letting the parameters in \cite[Eq. 8-73]{balanis2016antenna} be $h= d_{pq,v}$, $d=d_{n_1n_{1}^{\prime},h}$, and $l=d_l$, respectively. For $p\ne q, n_1=n_{1}^{\prime}$, $R_{r,n_1n_{1}^{\prime}}^{pq}$ and $X_{r,n_1n_{1}^{\prime}}^{pq}$ can be computed for the ``collinear" scenario in \cite[Eq. 8-72]{balanis2016antenna} by letting the variables in \cite[Eq. 8-72]{balanis2016antenna} be $h= d_{pq,v}$, $d=d_{n_1n_{1}^{\prime},h}$, and $l=d_l$, respectively. In summary, we can model $\mathbf{Z}_{r,C}$ based on the above terms. Similarly, we can also easily model $\mathbf{Z}_{t,C}$ based on the physical configurations of the transmitter.

Then, we focus on the computation of EDoF performance with the mutual coupling property. For the scalar Green's function-based channel, the channel matrix with the involvement of the mutual coupling can be denoted as
$\tilde{\mathbf{H}}_S=\mathbf{Z}_r\mathbf{H}_S\mathbf{Z}_t$.
Correspondingly, based on the analytical framework in Section~\ref{UPA_EDOF_Ana}, we can compute the EDoF performance with mutual coupling as $\tilde{\varepsilon}_S={\mathrm{tr}^2( \tilde{\mathbf{R}}_S )}/{\| \tilde{\mathbf{R}}_S \| _{\mathrm{F}}^{2}}$, where $\tilde{\mathbf{R}}_S=\tilde{\mathbf{H}}_{S}^{H}\tilde{\mathbf{H}}_S$. As for the scenario with the dyadic Green's function-based channel, we assume that the mutual coupling matrices for all polarization directions are the same. Thus, with the mutual coupling, the channel matrix between the $p$ polarization direction of the receiver and the $q$ polarization direction of the transmitter can be denoted as $\tilde{\mathbf{H}}^{pq}=\mathbf{Z}_r\mathbf{H}^{pq}\mathbf{Z}_t$. Moreover, exploiting a similar representation method as in \eqref{HD_Dis}, we can derive the dyadic Green's function-based channel matrix with the mutual coupling as $\tilde{\mathbf{H}}_D$. The EDoF performance for the dyadic Green's function-based channel with the mutual coupling can be computed as $\tilde{\varepsilon}_D={\mathrm{tr}^2( \tilde{\mathbf{R}}_D )}/{\| \tilde{\mathbf{R}}_D \| _{\mathrm{F}}^{2}}$ with $\tilde{\mathbf{R}}_D=\tilde{\mathbf{H}}_{D}^{H}\tilde{\mathbf{H}}_D$.

\section{Numerical Results}\label{num}
In this paper, we study the EDoF performance for XL-MIMO systems over the scalar and dyadic Green's function-based channels. Unless otherwise specified, for the UPA-based and 2D CAP plane-based XL-MIMO system, we consider the square plane, that is $L_{t,V}=L_{t,H}$ and $L_{r,V}=L_{r,H}$. Besides, we consider that the numbers of antennas per side of the UPA are equal, that is $M_{V}=M_{H}$ and $N_{V}=N_{H}$. The carrier frequency considered in this paper is $30\,\text{GHz}$. Note that, to fully capture the potential EDoF performance limits of considered systems, we consider the mutual coupling effect only in Fig.~\ref{figmutual}.

\begin{figure}[t]
\centering
\includegraphics[scale=0.43]{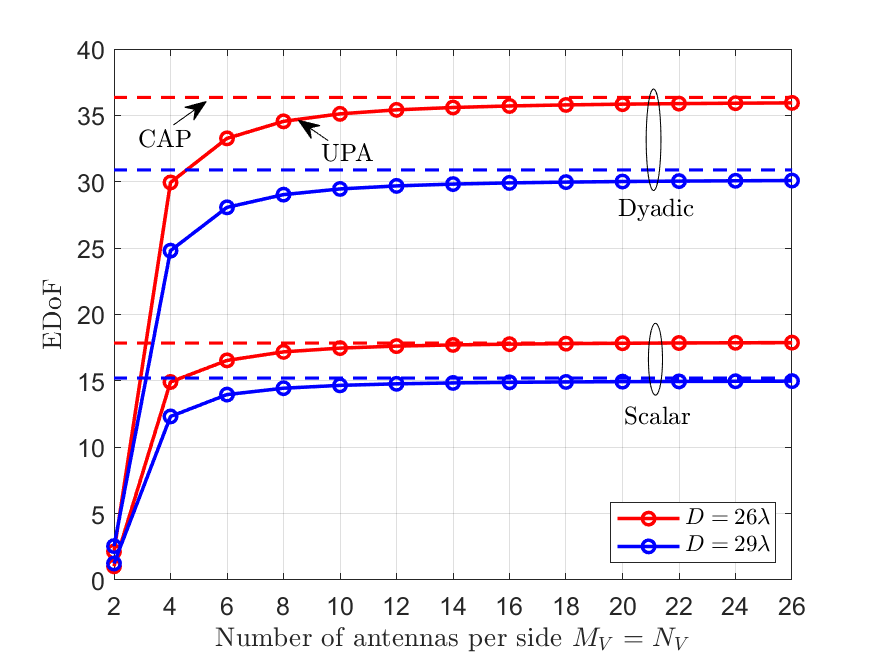}
\vspace{-0.3cm}
\caption{EDoF performance for UPA-based with infinitely thin dipoles and 2D CAP plane-based XL-MIMO system with $L_{t,V}=L_{r,V}=10\lambda$ over the dyadic and scalar Green's function-based channels.\label{1}}
\vspace{-0.4cm}
\end{figure}

\begin{figure}[t]
\centering
\includegraphics[scale=0.43]{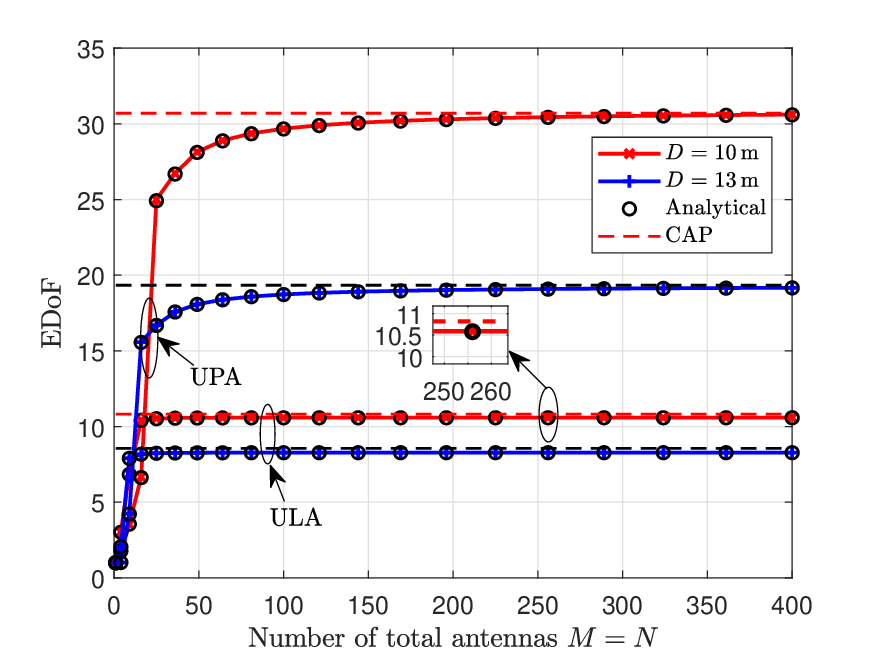}
\vspace{-0.3cm}
\caption{EDoF performance for the square UPA-based XL-MIMO system with $L_{t,V}=L_{r,V}=\sqrt{2}/2\,  \mathrm{m}$ and $M_{H,V}=N_{H,V}
$, and the ULA-based XL-MIMO system with $L_{t}=L_{r}=1\, \mathrm{m}$ and $M=N$ against the total number of antennas $M=N$ over the scalar Green's function-based channel. Each antenna element is an infinitely thin dipole. \label{6}}
\vspace{-0.4cm}
\end{figure}

\begin{figure}[t]
\centering
\includegraphics[scale=0.43]{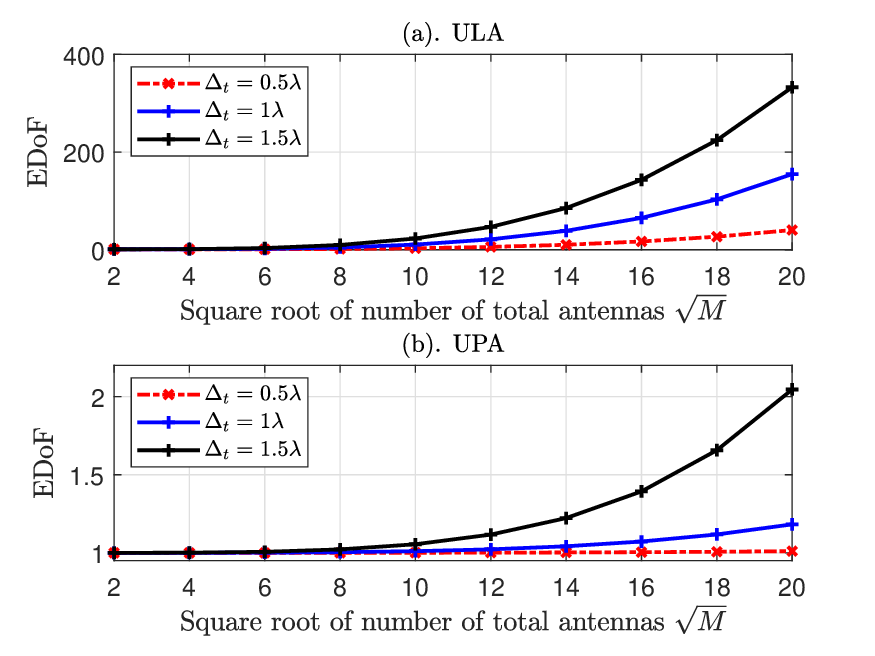}
\vspace{-0.3cm}
\caption{EDoF performance for the UPA-based XL-MIMO system with $M_{H,V}=N_{H,V}=\sqrt{M}$ and ULA-based XL-MIMO system with $M=N$ against $\sqrt{M}$ over different antenna spacing. Each antenna element is an infinitely thin dipole. \label{7}}
\vspace{-0.4cm}
\end{figure}

\begin{figure}[t]
\centering
\includegraphics[scale=0.43]{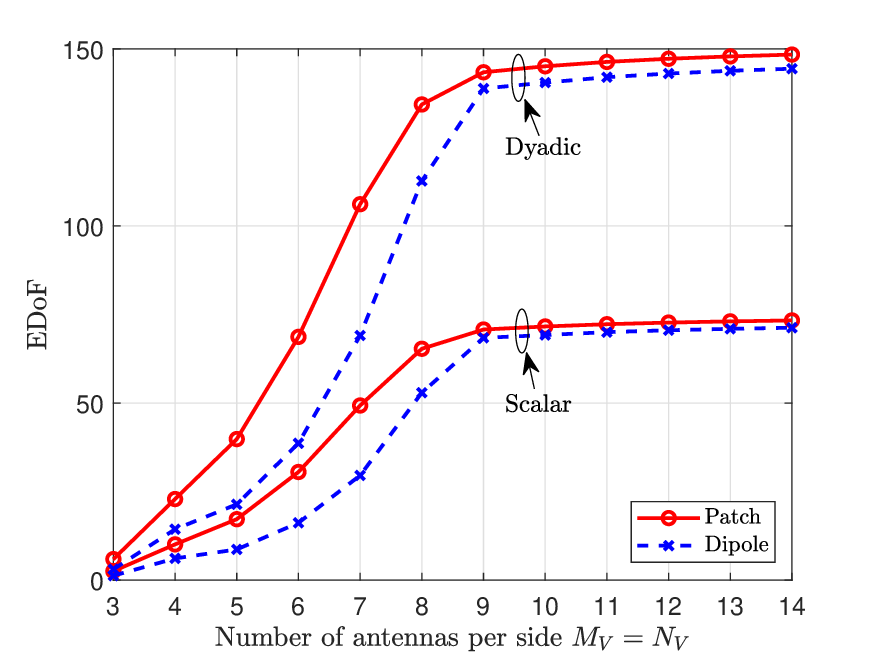}
\vspace{-0.3cm}
\caption{EDoF performance for the UPA-based XL-MIMO system with infinitely thin dipoles or patch antenna elements against the number of antennas per side $M_V=N_V$ with $L_{t,V}=L_{r,V}=10\lambda$, $D=10\lambda$, and $A_{r,\left\{ H,V \right\}}=A_{t,\left\{ H,V \right\}}=\lambda /2$
over both the dyadic and scalar Green's function-based channels.\label{5}}
\vspace{-0.4cm}
\end{figure}

\begin{figure}[t]
\centering
\includegraphics[scale=0.43]{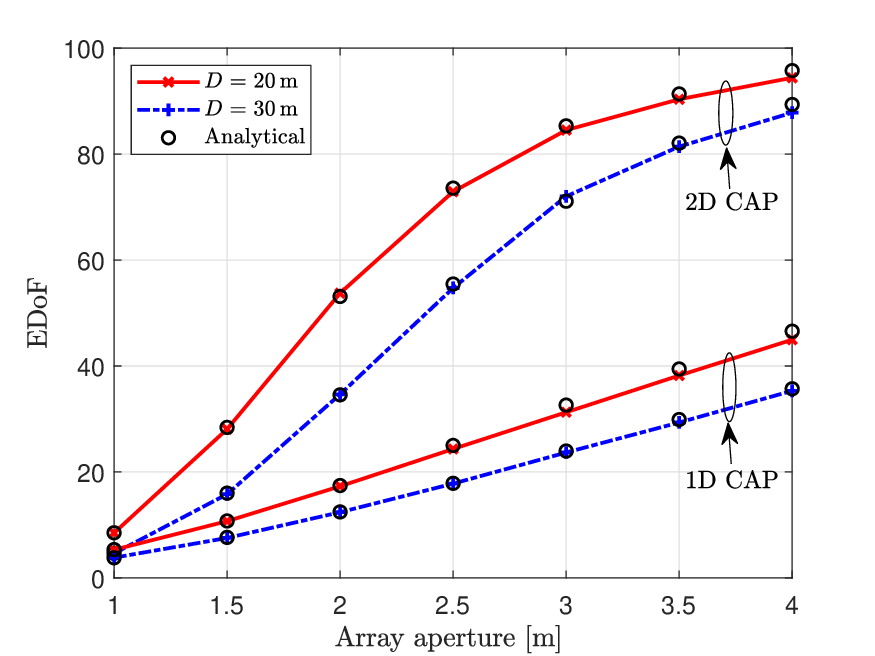}
\vspace{-0.3cm}
\caption{EDoF performance for 2D CAP plane-based and 1D CAP line segment-based XL-MIMO systems with the same array aperture size over the scalar Green's function-based channel over different values of the array aperture size.\label{8}}
\vspace{-0.4cm}
\end{figure}

\begin{figure}[t]
\centering
\includegraphics[scale=0.43]{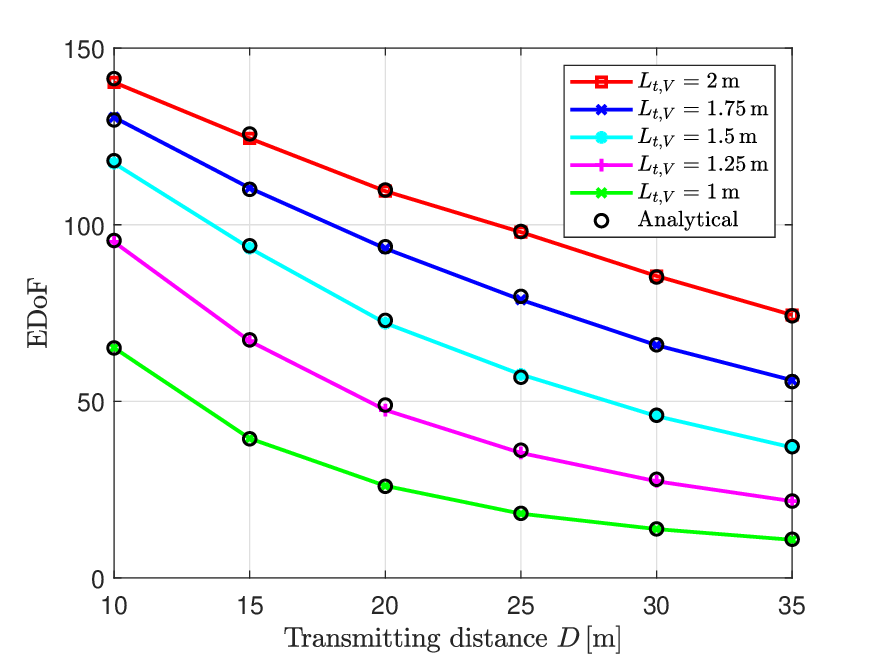}
\vspace{-0.3cm}
\caption{EDoF performance for the 2D CAP square plane-based XL-MIMO system against the transmitting distance $D$ with different values of $L_{t,V}=L_{r,V}$ over the scalar Green's function-based channel.
 \label{3}}
\vspace{-0.4cm}
\end{figure}

\begin{figure}[t]
\centering
\includegraphics[scale=0.43]{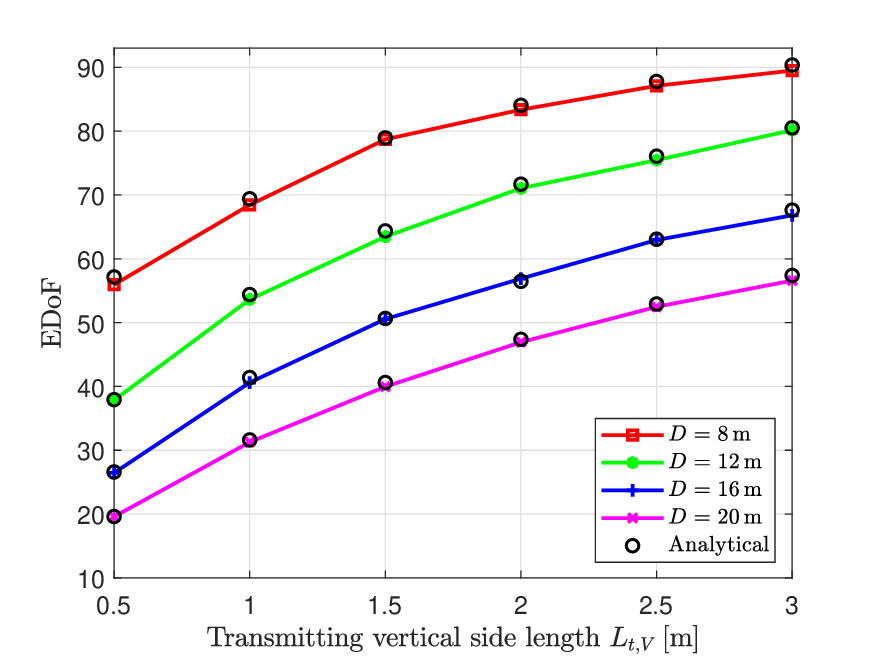}
\vspace{-0.3cm}
\caption{EDoF performance for the 2D CAP rectangle plane-based XL-MIMO system against the transmitting vertical side-length $L_{t,V}$ with $L_{t,H}=L_{r,H}= 1\,\mathrm{m}$ and $L_{r,V}= 1.5\,  \mathrm{m}$ over the scalar Green's function-based channel.\label{4}}
\vspace{-0.4cm}
\end{figure}

\begin{figure}[t]
\centering
\includegraphics[scale=0.43]{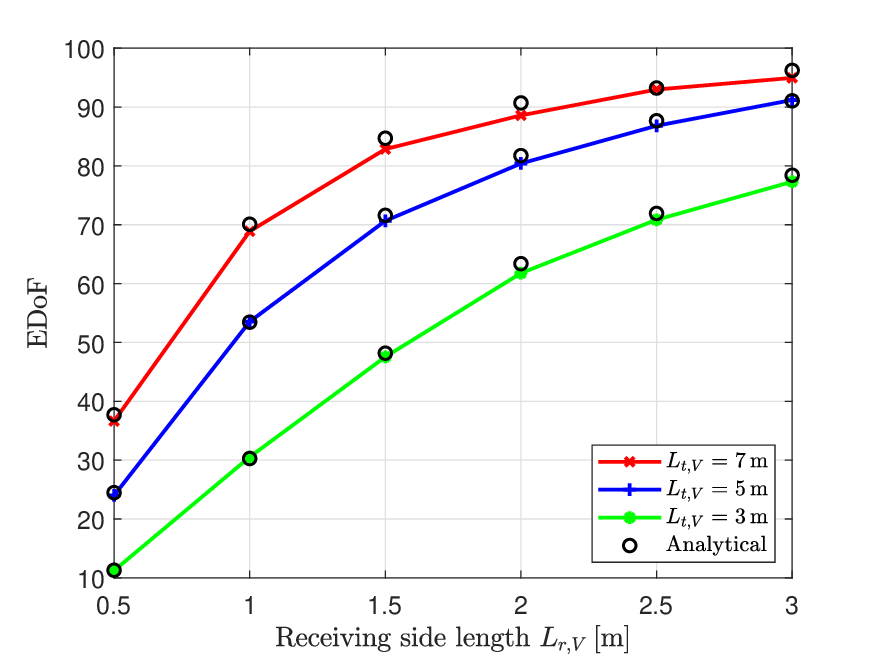}
\vspace{-0.3cm}
\caption{EDoF performance for the 2D CAP square plane-based XL-MIMO system against the receiving side-length $L_{r,V}=L_{r,H}$  over the scalar Green's function-based channel with $D=50\,  \mathrm{m}$. The analytical markers are generated based on the approximate closed-form results in Remark~\ref{remark4}. \label{Appr_closed}}
\vspace{-0.4cm}
\end{figure}

\begin{figure}[t]
\centering
\includegraphics[scale=0.43]{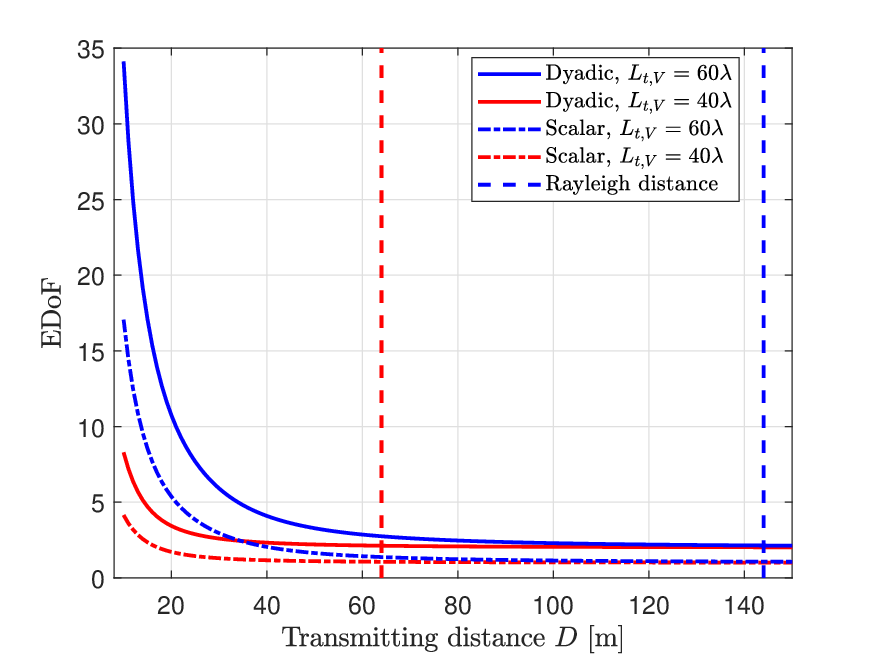}
\vspace{-0.3cm}
\caption{EDoF performance for the square UPA-based XL-MIMO system with infinitely thin dipoles against the transmitting distance $D$ with different values of $L_{t,V}=L_{r,V}$.\label{nearfar}}
\vspace{-0.4cm}
\end{figure}

\begin{figure}[t]
\centering
\includegraphics[scale=0.43]{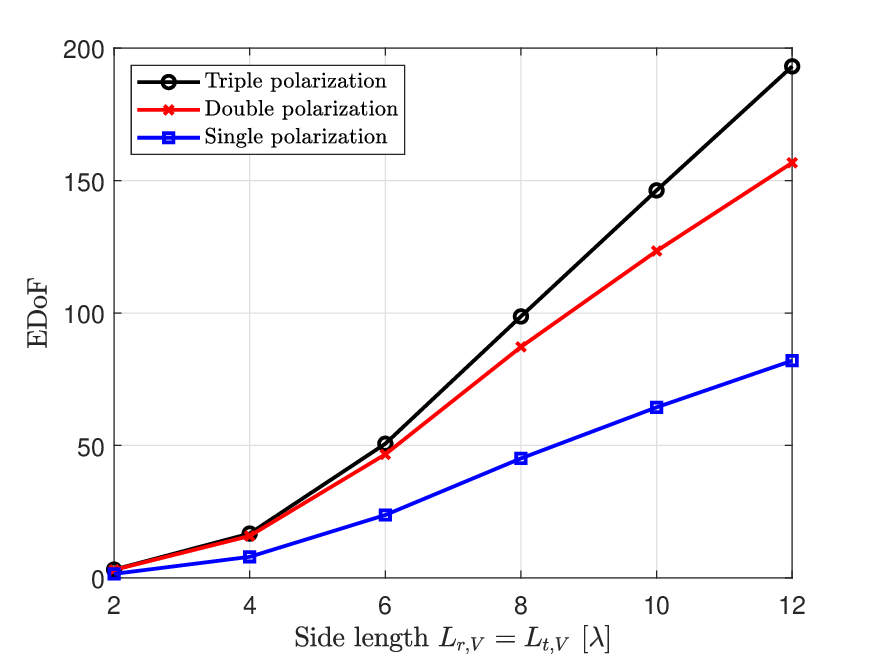}
\vspace{-0.3cm}
\caption{EDoF performance for 2D CAP square plane-based XL-MIMO system against the side-length $L_{t,V}=L_{r,V}$ over Green's function-based channels with different numbers of polarization with $D=6\lambda$. \label{2}}
\vspace{-0.4cm}
\end{figure}

\begin{figure}[t]
\centering
\includegraphics[scale=0.43]{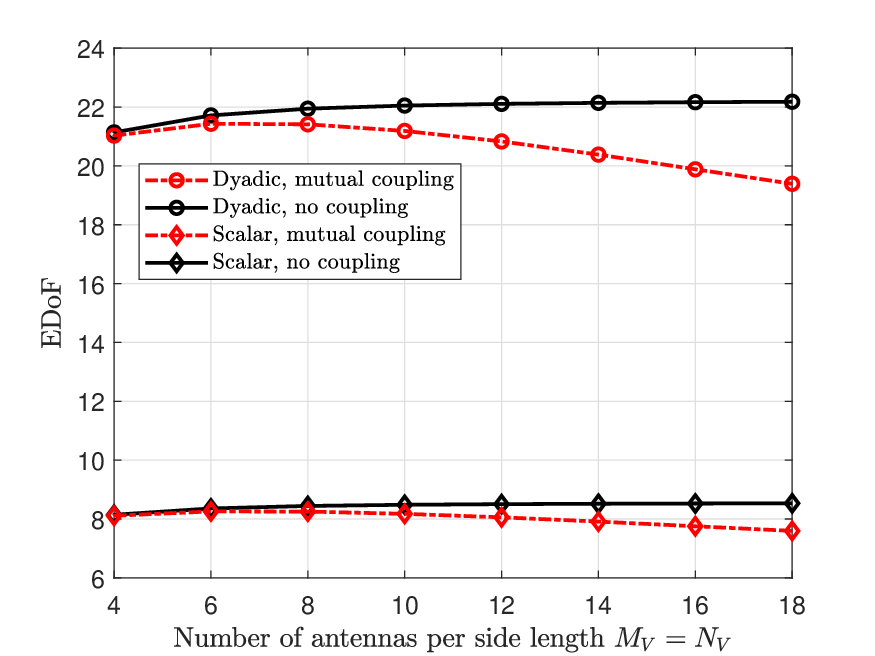}
\vspace{-0.3cm}
\caption{Impact of the mutual coupling on the EDoF performance for UPA-based with infinitely thin dipoles with $L_{t,V}=L_{r,V}=2\lambda$ and $D=1\lambda$. For the parameters in Section~\ref{mutual}, we have $Z_L=50\,\Omega$, $\eta =120\,\pi $, $\gamma _0=0.577$, $d_l=0.1\lambda $, and, $a=10^{-5}\lambda $.\label{figmutual}}
\vspace{-0.4cm}
\end{figure}

\subsection{Performance Comparison Among Different Designs}
In this part, we comprehensively compare and analyze the EDoF performance for all considered XL-MIMO hardware designs. 
In Fig.~\ref{1}, we compare the EDoF performance between the UPA-based XL-MIMO system with infinitely thin dipoles and the 2D CAP plane-based XL-MIMO system over both the dyadic and scalar Green's function-based channels. We can observe that as the number of antennas of the UPA increases, the EDoF performance for the UPA-based system approaches that of the CAP-based system over both the dyadic and scalar Green's function-based channels. This observation indicates that the physical size of the XL-MIMO system imposes restrictive constraints on the EDoF performance. It is vital to observe that increasing the number of antennas in a fixed physical size may not always benefit the EDoF performance, but lead to an approximately saturated value. Besides, the dyadic Green's function-based channel can significantly enhance the EDoF performance compared with that of the scalar Green's function channel due to the consideration of full polarization, e.g. about $104\%$ EDoF improvement for $D=26\lambda$, highlighting the importance of utilizing the dyadic Green's function-based channel to fully exploit the EDoF performance of XL-MIMO.

In Fig.~\ref{6} and Fig.~\ref{7}, we compare the EDoF performance between the UPA-based XL-MIMO system with infinitely thin dipoles and the ULA-based XL-MIMO system with infinitely thin dipoles over the scalar Green's function-based channel. To ensure fairness in the comparison, it is important to establish appropriate criteria \cite{ZheSurvey}. The first potential criteria is to compare the EDoF performance for the UPA-based and ULA-based systems with the same array aperture. For the ULA-based system utilizing a 1D array, the array aperture for the transmitter/receiver is the respective side-length. For the UPA-based system, which employs a 2D array, the array aperture for the transmitter/receiver is their respective length of the diagonal of the array. As observed in Fig. \ref{6}, with the same array aperture size, the UPA-based system can achieve remarkably superior EDoF performance compared with the ULA-based system. Indeed, with the same array aperture size, 2D spatial resources can be effectively utilized to enhance the EDoF performance in the UPA-based system. We also validate the EDoF closed-form results for the UPA-based and ULA-based XL-MIMO systems in Theorem~\ref{UPAclosed} and Theorem~\ref{TheoremULAclosed}, respectively. Moreover, as the number of antennas increases, the EDOF performance for both the UPA-based and ULA-based systems approaches to that of respective CAP-based systems.

Another comparison criteria is to consider the UPA-based and ULA-based with the same number of antennas. In Fig.~\ref{7}, we compare the EDoF performance for the UPA-based and ULA-based systems with the same total numbers of antennas, i.e., same $M=N$, over different values of antenna spacing. We assume that $\Delta _{t,\left\{ H,V \right\}}=\Delta _{r,\left\{ H,V \right\}}=\Delta _{t}=\Delta _{r}$. Under this setting, the physical size of the transceiver can be adjusted according to the variations in the number of antennas. We observe from Fig.~\ref{7} that the ULA-based system can achieve significantly superior EDoF performance compared with the UPA-based system. This phenomenon is attributed to the much larger array aperture size for the ULA-based system compared with that of the UPA-based system, e.g. $6 \,\mathrm{m}$ and $0.42 \,\mathrm{m}$ array aperture for the ULA-based and UPA-based system with $\Delta _t=1.5\lambda$, respectively. Due to the 2D antenna deployment characteristic, the array aperture for the UPA-based system remains relatively small, even with a large number of antennas, such as $M=400$.

Fig.~\ref{5} compares the EDoF performance for the UPA-based system with infinitely thin dipoles and that of the UPA-based system with patch antennas. Note that the design with patch antennas can achieve better EDoF performance than the design with infinitely thin dipoles when the antennas are deployed in a non-dense manner, i.e., when $M_V=N_V$ are small. When the antennas are deployed non-densely, the EDoF performance for the UPA-based has not yet reached its saturated values. In such a case, patch antennas with a certain element size can efficiently capture EM waves across their element regions, resulting in better EDoF performance compared to infinitely thin dipoles. For instance, for the case with $M_V=N_V=7$, the scenario with patch antennas can achieve $53.7\%$ and $66.5\%$ EDoF performance improvement over the dyadic and scalar Green's function-based channels, respectively, compared with the scenario with infinitely thin dipoles. However, as the number of antennas further increases, the performance gaps between the patch antenna scenario and the infinitely thin dipole scenario reduce, since both the scenarios with patch antennas and infinitely thin dipoles would eventually reach approximately equal saturated EDoF values, limited by the physical sizes of the transmitter and the receiver.

As can be observed from the above, as the number of antennas increases, the EDoF performance for discrete XL-MIMO designs, i.e., UPA-based and ULA-based designs, approaches that of their respective continuous XL-MIMO designs, i.e., 2D CAP plane-based and 1D CAP line segment-based designs, respectively. Thus, it is interesting to compare the EDoF performance between 2D CAP plane-based and 1D CAP line segment-based systems to bring a comprehensive vision of the EDoF performance of all XL-MIMO designs. It can be observed from Fig.~\ref{8} that with the same aperture size, the 2D CAP plane-based system can achieve higher EDoF performance compared to the 1D CAP line segment-based system, e.g. $109.9\%$ EDoF performance enhancement with the array aperture of $4\,\mathrm{m}$ and $D=20\,\mathrm{m}$. Moreover, we can validate the EDoF closed-form results for both the 2D CAP system and 1D CAP system derived in Theorem~\ref{CAPclosed} and Theorem~\ref{Theorem1DCAPclosed}, respectively. Based on the above findings, we demonstrate that the EDoF performance is primarily constrained and determined by the physical size (array aperture) rather than the number of antennas. From the goal of higher EDoF, the plane-based system or the line segment-based system is advocated when designing the system with a certain length of array aperture or with a certain number of antennas.

\subsection{Verification of Closed-Form Results in Theorem~\ref{CAPclosed}}
In this subsection, we provide various simulation results to validate the derived closed-form results in Theorem~\ref{CAPclosed}.
Fig.~\ref{3} investigates the EDoF performance for the 2D CAP square plane-based XL-MIMO system against the transmitting distance over the scalar Green's function-based channel. We observe that the EDoF performance benefits from the increase of the side-length or the decrease of the transmitting distance. More importantly, it can be found that markers ``$\circ$" generated by the EDoF closed-form results as described in Theorem~\ref{CAPclosed} match well with the curves generated by the simulation results, which validates our derived EDoF closed-form results for the 2D CAP plane-based XL-MIMO system in Theorem~\ref{CAPclosed} over different values of the side-lengths and transmitting distances. 

In Fig.~\ref{4}, we provide further validation of the EDoF closed-form results in Theorem~\ref{CAPclosed}. We consider the 2D CAP rectangle plane-based XL-MIMO system with transmitting rectangle planes of different shapes over the scalar Green's function-based channel. As observed, the EDoF closed-form results, denoted by ``$\circ$", match well with the simulation results denoted by the curves, confirming the accuracy of our derived EDoF closed-form results in rectangle plane-based scenarios with varying side-lengths. Moreover, with the increase of $L_{t,V}$, the EDoF performance increases with diminishing returns, due to the performance constraint introduced by the limited physical size of the receiver, e.g. $22.3\%$ and $2.7\%$ EDoF improvement for $L_{t,V}=1\,\mathrm{m}$ compared with $L_{t,V}=0.5\,\mathrm{m}$ and  $L_{t,V}=3\,\mathrm{m}$ compared with $L_{t,V}=2.5\,\mathrm{m}$, respectively, for $D=8\,\mathrm{m}$. Furthermore, in Fig.~\ref{Appr_closed}, we validate the accuracy of the approximate closed-form results in Remark~\ref{remark4}, which are derived based on the prerequisite that the physical size of the transmitting plane is much larger than that of the receiving plane. As observed, although the closed-form results in Remark~\ref{remark4} are approximate, the markers ``$\circ$" generated by them match well with the curves generated by the simulation results, especially in scenarios where $L_{t,V}$ is much larger than $L_{r,V}$.

\subsection{Impacts of the EM Properties in XL-MIMO}
In this subsection, we discuss the impacts of the vital EM properties in XL-MIMO, e.g. the near-field spherical wave, polarization, and mutual coupling, on the EDoF performance. In Fig.~\ref{nearfar}, we show the EDoF performance against the transmitting distance $D$ for the UPA-based XL-MIMO system with infinitely thin dipoles. More importantly, we highlight the Rayleigh distance, which is considered to be the most representative boundary dividing the near-field and far-field regions \cite{ZheSurvey}. As observed, the involvement of Green's functions-based near-field scenarios significantly improves the EDoF performance compared with those of the planar wave-based far-field scenarios. Moreover, the EDoF in far-field scenarios utilizing scalar and dyadic Green's functions-based channel model are $1$ and $2$, respectively, validating the assertions made in Remark~\ref{farnear}.

Fig.~\ref{2} evaluates the effects of polarization in channels on the EDoF performance. We consider the EDoF performance for the 2D CAP square plane-based XL-MIMO system as a function of the side-length over channels with different numbers of polarization. We can observe that the triple and double polarized channels can enhance the EDoF performance compared with that of the single polarized channel, i.e., scalar Green's function-based channel. Furthermore, with the increase of the side-length, the EDoF performance improvement for the triple polarized channel compared with that of the double and single polarized channels becomes larger, e.g. $23.2\%$ and $135.5\%$ improvement for $L_{t,V}=12\lambda$ as well as $8.6\%$ and $113.4\%$ improvement for $L_{t,V}=6\lambda$, respectively. More importantly, we observe that when the side-length is small, the EDoF performance of the triple polarized channel closely approaches the EDoF performance of the double polarized channel. This phenomenon is due to the fact that when the transmitting distance is comparable to the physical size of the system, the effect of $z$ polarization vanishes. In other words, for an XL-MIMO system with a fixed physical size, the effect of $z$ polarization would diminish and eventually vanish with the increase of the transmitting distance $D$.

In Fig.~\ref{figmutual}, we study the impact of the mutual coupling effect on EDoF performance. As observed, with the involvement of mutual coupling, the EDoF performance in both the dyadic and scalar Green's function-based scenarios first reaches maximum values and then degrades as the number of antennas increases. This is due to the mutual coupling effect becoming more severe with the decrease in antenna spacing, induced by the increase in the number of antennas in the invariant array aperture. Besides, the performance gap between the EDoF performance with and without the involvement of mutual coupling becomes larger with the increase of the number of antennas. For instance, there are about $12.6\%$ and $11\%$ performance gaps between the EDoF performance with and without mutual coupling in the dyadic and scalar Green's functions-based scenarios, respectively. Furthermore, it is vital to select the proper number of antennas to achieve the optimal EDoF performance due to the existence of the practical mutual coupling effect.

\begin{figure*}[t]
{{\begin{align}\tag{44} \label{UPA_Closed_Distance} 
d_{nm_1}&=\sqrt{D^2+( -\frac{L_{r,H}}{2}+t\left( n \right) \Delta _{r,H}+\frac{L_{t,H}}{2}-i\left( m \right) \Delta _{t,H} ) ^2+( -\frac{L_{r,V}}{2}+k\left( n \right) \Delta _{r,V}+\frac{L_{t,V}}{2}-j\left( m \right) \Delta _{t,V} ) ^2}\\
&\approx D+\small{\frac{( -\frac{L_{r,H}}{2}+t\left( n \right) \Delta _{r,H}+\frac{L_{t,H}}{2}-i\left( m \right) \Delta _{t,H} ) ^2+( -\frac{L_{r,V}}{2}+k\left( n \right) \Delta _{r,V}+\frac{L_{t,V}}{2}-j\left( m \right) \Delta _{t,V} ) ^2}{2D}}\notag
\end{align}}
\hrulefill
\vspace*{-0.6cm}
%\vspace*{3pt}
}\end{figure*}

\begin{figure*}[t]
{{\begin{align}\tag{45} \label{probability_theory_step}
\gamma &=\int_{S_R}{\int_{S_T}{|G(\mathbf{r}_r,\mathbf{r}_t)|^2d\mathbf{r}_td\mathbf{r}_r}}=\mu _0\int_{-\frac{L_{t,H}}{2}}^{\frac{L_{t,H}}{2}}{\int_{-\frac{L_{t,V}}{2}}^{\frac{L_{t,V}}{2}}{\int_{-\frac{L_{r,H}}{2}}^{\frac{L_{r,H}}{2}}{\int_{-\frac{L_{r,V}}{2}}^{\frac{L_{r,V}}{2}}{\frac{dx_rdy_rdx_tdy_t}{D^2+\left( x_r-x_t \right) ^2+\left( y_r-y_t \right) ^2}}}}}\\
&=\mu _0L_{t,H}L_{r,H}L_{t,V}L_{r,V}\int_0^{\frac{L_{t,H}+L_{r,H}}{2}}{\int_0^{\frac{L_{t,V}+L_{r,V}}{2}}{\frac{f\left( x \right) g\left( y \right) dxdy}{D^2+x^2+y^2}}} \notag
\end{align}}
\hrulefill
\vspace*{-0.6cm}
%\vspace*{3pt}
}\end{figure*}

\begin{figure*}[t]
{{\begin{align}\tag{51} \label{r12} 
&\gamma _{12}=\frac{L_{t,V}+L_{r,V}}{L_{t,V}L_{r,V}}\frac{1}{\sqrt{D^2+x^2}}\left. \arctan{\frac{y}{\sqrt{D^2+x^2}}} \right|_{\frac{\left| L_{t,V}-L_{r,V} \right|}{2}}^{\frac{L_{t,V}+L_{r,V}}{2}}-\frac{1}{L_{t,V}L_{r,V}}\left. \ln \left( y^2+D^2+x^2 \right) \right|_{\frac{\left| L_{t,V}-L_{r,V} \right|}{2}}^{\frac{L_{t,V}+L_{r,V}}{2}}\\
&\!=\!\frac{L_{t,V}+L_{r,V}}{L_{t,V}L_{r,V}}\frac{1}{\sqrt{D^2+x^2}}\!\left( \arctan{\frac{L_{t,V}+L_{r,V}}{2\sqrt{D^2+x^2}}}\!-\!\arctan{\frac{\left| L_{t,V}-L_{r,V} \right|}{2\sqrt{D^2+x^2}}} \right)\! \!-\!\frac{1}{L_{t,V}L_{r,V}}\ln \left( \frac{\mu_1+4x^2 }{\mu_2+4x^2} \right).\notag
\end{align}}
\hrulefill
\vspace*{-0.6cm}
%\vspace*{3pt}
}\end{figure*}

\begin{figure*}[t]
{{\begin{align} \tag{53}\label{r_final}
\gamma =\frac{1}{\left( 4\pi \right) ^2}L_{t,H}L_{r,H}\int_0^{\frac{\left| L_{t,H}-L_{r,H} \right|}{2}}{\gamma _1\frac{2}{L_{H,\max}}dx}+\frac{1}{\left( 4\pi \right) ^2}L_{t,H}L_{r,H}\int_{\frac{\left| L_{t,H}-L_{r,H} \right|}{2}}^{\frac{L_{t,H}+L_{r,H}}{2}}{\gamma _1\frac{L_{t,H}+L_{r,H}-2x}{L_{t,H}L_{r,H}}dx}
\end{align}}
\hrulefill
\vspace*{-0.6cm}
%\vspace*{3pt}
}\end{figure*}

\section{Conclusions}\label{con}

In this paper, we presented a comprehensive investigation of the EDoF performance analysis framework for near-field XL-MIMO systems. We studied both the discrete XL-MIMO designs, i.e., UPA-based XL-MIMO designs with infinitely thin dipoles or patch antennas and ULA-based XL-MIMO design, and the continuous XL-MIMO designs, i.e. 2D CAP plane-based XL-MIMO design and 1D CAP line segment-based XL-MIMO design. Notably, scalar and dyadic Green's function-based near-field channel models were considered. When applying the scalar channels, we derived the closed-form EDoF expressions. In numerical results, we thoroughly evaluated and compared the EDoF performance for all studied scenarios. The accuracy of the proposed closed-form results was validated through extensive simulation results. More significantly, we observed that the consideration of multiple polarization and the increase of array aperture could benefit the EDoF performance. In particular, the EDoF is constrained by the array aperture size and thus shows saturated values as the number of antennas increases.

\begin{appendices}
\section{Useful Results}
In this appendix, we provide some useful basics, which will be applied for the computation of EDoF closed-form expressions later. For a real-valued number $x$, by applying the first-order Taylor expansion, we have
\begin{equation}\label{Taylor}
\sqrt{x+1}\approx 1+\frac{x}{2}.
\end{equation}
Besides, we have
\begin{equation}\label{arctan}
\arctan{x}\approx x.
\end{equation}
Moreover, for two variables $x$ and $y$ approaching $0$, we have
\begin{equation}\label{square}
x^2+y^2\approx \frac{\left( x+y \right)}{2}^2
\end{equation}

Then, we list the indefinite integral expressions used in the following. With a constant $a$, we have

\begin{align}
\int{\frac{x}{x^2+a^2}dx}&=\frac{1}{2}\ln \left( x^2+a^2 \right) , \label{integral1} \\
\int{\frac{1}{x^2+a^2}dx}&=\frac{1}{a}\arctan{\frac{x}{a}}\,\,a>0, \label{integral2} \\
\int{\ln \left( a^2+x^2 \right) dx}&=x\ln \left( a^2+x^2 \right) -2x+2a\arctan{\frac{x}{a}}, \label{integral3}\\
\int{x\ln \left( a^2+x^2 \right) dx}&=\frac{x^2+a^2}{2}\ln \left( a^2+x^2 \right) -\frac{1}{2}x^2.\label{integral4}
\end{align}
Moreover, with constants $a$ and $b$, we have
\begin{align}
\int{\frac{a}{\left( b+y^2 \right) ^2}}dy&=\frac{ay}{2b\left( y^2+b \right)}+\frac{a}{2b\sqrt{b}}\arctan{\left( \frac{y}{\sqrt{b}} \right)} ,\label{integral6}\\
\int{\frac{ay}{\left( b+y^2 \right) ^2}dy}&=-\frac{a}{2\left( b+y^2 \right)}.\label{integral7}
\end{align}

\begin{figure*}[t]
{{\begin{align} \tag{54}\label{Kernel_expand}
&K(\mathbf{r}_t,\mathbf{r}_{t^{\prime}})%=\int_{x_r}{\int_{y_r}{\frac{\exp ( j\kappa _0\sqrt{( x_t-x_r ) ^2+( y_t-y_r) ^2+D^2}) \exp ( -j\kappa _0\sqrt{( x_{t^{\prime}}-x_r) ^2+( y_{t^{\prime}}-y_r) ^2+D^2} )}{\left( 4\pi \right) ^2\sqrt{[ ( x_t-x_r ) ^2+( y_t-y_r) ^2+D^2 ] [ ( x_{t^{\prime}}-x_r) ^2+( y_{t^{\prime}}-y_r ) ^2+D^2 ]}}dx_r}}dy_r\\ \notag
\overset{\left( a \right)}{\approx}\int_{x_r}{\int_{y_r}{\frac{\exp ( j\kappa _0\sqrt{( x_t-x_r ) ^2+( y_t-y_r ) ^2+D^2} ) \exp ( -j\kappa _0\sqrt{( x_t-x_r ) ^2+( y_t-y_r ) ^2+D^2} )}{\left( 4\pi \right) ^2[ D^2+( \frac{y_t+y_{t^{\prime}}}{2} ) ^2+( \frac{x_t+x_{t^{\prime}}}{2} ) ^2 ]}dx_r}}dy_r\\
&\overset{\left( b \right)}{\approx}\frac{L_{r,H}L_{r,V}\exp ( j\kappa _0\sqrt{( x_t-x_r ) ^2+( y_t-y_r ) ^2+D^2} ) \exp ( -j\kappa _0\sqrt{( x_t-x_r ) ^2+( y_t-y_r ) ^2+D^2} )}{\left( 4\pi \right) ^2[ D^2+( \frac{y_t+y_{t^{\prime}}}{2} ) ^2+( \frac{x_t+x_{t^{\prime}}}{2} ) ^2 ]}\notag
\end{align}}
\hrulefill
\vspace*{-0.6cm}
%\vspace*{3pt}
}\end{figure*}

%\newcounter{mytempeqncnt12}
%\begin{figure*}[t]
%\normalsize
%\setcounter{mytempeqncnt12}{\value{equation}}
%\setcounter{equation}{2}
%\begin{align} \label{K_dis}\notag
%\int_{S_T}{\int_{S_T}{\left| K(\mathbf{r}_t,\mathbf{r}_{t^{\prime}}) \right|^2}d}\mathbf{r}_td\mathbf{r}_{t^{\prime}}=\sum_{j=1}^{M_s}{\sum_{w=1}^{M_s}{\sum_{i=1}^{N_s}{\sum_{k=1}^{N_s}{( \frac{L_{t,H}L_{r,H}}{M_sN_s})}^2}}}\int_{-\frac{L_{t,V}}{2}}^{\frac{L_{t,V}}{2}}{\int_{-\frac{L_{t,V}}{2}}^{\frac{L_{t,V}}{2}}{\frac{\left( \mu _0L_{r,V} \right) ^2}{( d_{kj}d_{kw}^{\prime}+( \frac{y_t+y_{t^{\prime}}}{2} ) ^2 ) ( d_{ij}d_{iw}^{\prime}+( \frac{y_t+y_{t^{\prime}}}{2} ) ^2 )}dy_tdy_{t^{\prime}}}}
%\end{align}
%\setcounter{equation}{\value{mytempeqncnt12}}
%\hrulefill
%\vspace*{-0.4cm}
%\end{figure*}

\begin{figure*}[t]
{{\begin{align}\tag{55} \label{d_approx}
&\sqrt{[ D^2+( x_t-x_r ) ^2+( y_t-y_r ) ^2 ] [ D^2+( x_{t^{\prime}}-x_r ) ^2+( y_{t^{\prime}}-y_r ) ^2 ]}\\ \notag
&=D^2\sqrt{\begin{array}{l}
	1+\frac{\left( x_{t^{\prime}}-x_r \right) ^2}{D^2}+\frac{\left( x_t-x_r \right) ^2}{D^2}+\frac{\left( y_{t^{\prime}}-y_r \right) ^2}{D^2}+\frac{\left( y_t-y_r \right) ^2}{D^2}+\frac{\left( y_t-y_r \right) ^2\left( x_{t^{\prime}}-x_r \right) ^2}{D^4}\\
	+\frac{\left( y_t-y_r \right) ^2\left( y_{t^{\prime}}-y_r \right) ^2}{D^4}+\frac{\left( x_t-x_r \right) ^2\left( x_{t^{\prime}}-x_r \right) ^2}{D^4}+\frac{\left( x_t-x_r \right) ^2\left( y_{t^{\prime}}-y_r \right) ^2}{D^4}\\
\end{array}}\overset{\left( c \right)}{\approx}D^2\sqrt{1+\frac{\left( x_{t^{\prime}}+x_t \right) ^2}{2D^2}+\frac{\left( y_{t^{\prime}}+y_t \right) ^2}{2D^2}}\\
&\overset{\left( d \right)}{\approx}D^2+( \frac{x_t+x_{t^{\prime}}}{2} ) ^2+( \frac{y_t+y_{t^{\prime}}}{2} ) ^2 \notag
\end{align}}
\hrulefill
\vspace*{-0.6cm}
%\vspace*{3pt}
}\end{figure*}

\begin{figure*}[t]
{{\begin{align} \tag{56}\label{K_dis}
\int_{S_T}{\int_{S_T}{\left| K(\mathbf{r}_t,\mathbf{r}_{t^{\prime}}) \right|^2}d}\mathbf{r}_td\mathbf{r}_{t^{\prime}}=\underset{M_s\rightarrow \infty}{\lim}\underset{N_s\rightarrow \infty}{\lim}\varphi \times \underset{\zeta}{\underbrace{\int_{S_T}{\int_{S_T}{| \frac{L_{r,H}L_{r,V}}{\left( 4\pi \right) ^2[ D^2+( \frac{y_t+y_{t^{\prime}}}{2} ) ^2+( \frac{x_t+x_{t^{\prime}}}{2} ) ^2 ]} |^2}d}\mathbf{r}_td\mathbf{r}_{t^{\prime}}}}
\vspace*{-1cm}
\end{align}}
\hrulefill
\vspace*{-0.6cm}
%\vspace*{3pt}
}\end{figure*}

\vspace{-0.6cm}

\section{Proof of Theorem~\ref{UPAclosed}}\label{UPAClosedProof}

Note that $\left[ \mathbf{R}_S \right] _{m_1m_2}=\sum_{n=1}^N{\frac{\exp \left( j\kappa _0\left( d_{nm_1}-d_{nm_2} \right) \right)}{(4\pi )^2d_{nm_1}d_{nm_2}}}$, where $d_{nm_1}=\left| \mathbf{r}_{r,n}-\mathbf{r}_{t,m_1} \right|$ and $d_{nm_2}=\left| \mathbf{r}_{r,n}-\mathbf{r}_{t,m_2} \right|$ are the distances between the $n$-th receiving antenna and the ${m_1,m_2}$-th transmitting antenna, respectively. By applying the coordinates of the receiving and transmitting antennas defined in Section~\ref{SystemUPA} and the first-order Taylor expansion as \eqref{Taylor}, we can denote $d_{nm_1}$ as \eqref{UPA_Closed_Distance}. By substituting \eqref{UPA_Closed_Distance} into \eqref{EDoFterms_Scalar}, we can compute the EDoF for the UPA-based XL-MIMO system over the scalar channel in the closed-form as Theorem~\ref{UPAclosed}.

\vspace{-0.3cm}
\section{Proof Theorem~\ref{CAPclosed}}\label{gamma}
We provide the detailed proof steps to compute the EDoF for the 2D CAP plane-based XL-MIMO system over the scalar Green's function-based channel in a closed-form. Firstly, we start from the computation of $\gamma$. Note that we define
$\gamma=\int_{S_R}{\int_{S_T}{|G(\mathbf{r}_r,\mathbf{r}_t)|^2d\mathbf{r}_td\mathbf{r}_r}}$ as the channel power gain of the 2D CAP plane-based XL-MIMO system over the scalar Green's function-based channel. Based on the probability theory, we can denote $\gamma$ as \eqref{probability_theory_step}, where $f\left( x \right)$ and $g\left( y \right)$ are the probability density functions (PDFs) of $x\triangleq \left| x_t-x_r \right|\in [ 0,\frac{L_{t,H}+L_{r,H}}{2} ] $ and $y\triangleq \left| y_t-y_r \right|\in [ 0,\frac{L_{t,V}+L_{r,V}}{2} ] $, respectively. As for the computation of PDFs, we take $f\left( x \right)$ as an example. Based on the value ranges of $x_t$ and $x_r$, we can compute the probability distribution function $F\left( x \right)$ based on the area ratio between the aimed region and the total value region. For the scenario with $L_{t,H}\geqslant L_{r,H}$, when $0\leqslant x\leqslant \frac{L_{t,H}-L_{r,H}}{2}$, we have $F\left( x \right) =\frac{2x\cdot L_{r,H}}{L_{t,H}L_{r,H}}=\frac{2x}{L_{t,H}}$. When $\frac{L_{t,H}-L_{r,H}}{2}\leqslant x\leqslant \frac{L_{t,H}+L_{r,H}}{2}$, we have
$
F\left( x \right) =[L_{t,H}L_{r,H}-2\times \frac{1}{2}( \frac{L_{t,H}+L_{r,H}}{2}-x ) ^2]/{L_{t,H}L_{r,H}}=1-{( x-\frac{L_{t,H}+L_{r,H}}{2} ) ^2}/{L_{t,H}L_{r,H}}.
$
For the scenario with $L_{r,H}\geqslant L_{t,H}$, similarly, we can derive
\addtocounter{equation}{2}
\begin{equation}
F\left( x \right) =\begin{cases}
	\frac{2x}{L_{r,H}}, \, 0\leqslant x\leqslant \frac{L_{r,H}-L_{t,H}}{2}\\
	1-\frac{\left( x-\frac{L_{t,H}+L_{r,H}}{2} \right) ^2}{L_{t,H}L_{r,H}}, \, \frac{L_{r,H}-L_{t,H}}{2}\leqslant x\leqslant \frac{L_{t,H}+L_{r,H}}{2}.\\
\end{cases}
\end{equation}
In summary, we have $f\left( x \right) =F^{\prime}\left( x \right)$ as
\begin{equation}\label{fx}
f\left( x \right) =\begin{cases}
	\frac{2}{L_{H,\max}}\,\,, 0\leqslant x\leqslant \frac{\left| L_{t,H}-L_{r,H} \right|}{2}\\
	\frac{L_{t,H}+L_{r,H}-2x}{L_{t,H}L_{r,H}}, \,\,  \frac{\left| L_{t,H}-L_{r,H} \right|}{2}\leqslant x\leqslant \frac{L_{t,H}+L_{r,H}}{2}.\\
\end{cases}
\end{equation}

Similarly, we can derive $g\left( y \right)$ as
\begin{equation}\label{gy}
g\left( y \right) =\begin{cases}
	\frac{2}{L_{V,\max}}\,\,, 0\leqslant y\leqslant \frac{\left| L_{t,V}-L_{r,V} \right|}{2}\\
	\frac{L_{t,V}+L_{r,V}-2y}{L_{t,V}L_{r,V}},\,\,  \frac{\left| L_{t,V}-L_{r,V} \right|}{2}\leqslant y\leqslant \frac{L_{t,V}+L_{r,V}}{2},\\
\end{cases}
\end{equation}
where $L_{V,\max}=\max \left\{ L_{t,V},L_{r,V} \right\} $. Then, we define
\begin{equation}\label{r1}
\begin{aligned}
\gamma _1&=L_{t,V}L_{r,V}\int_0^{\frac{L_{t,V}+L_{r,V}}{2}}{\frac{g\left( y \right)}{D^2+x^2+y^2}dy}\\
&=L_{t,V}L_{r,V}\left( \gamma _{11}+\gamma _{12} \right),
\end{aligned}
\end{equation}
where $\gamma _{11}=\int_0^{\frac{\left| L_{t,V}-L_{r,V} \right|}{2}}{\frac{1}{D^2+x^2+y^2}\frac{2}{L_{V,\max}}dy}$ and $\gamma _{12}\!=\!\!\int_{\frac{\left| L_{t,V}-L_{r,V} \right|}{2}}^{\frac{L_{t,V}+L_{r,V}}{2}}{\frac{1}{D^2+x^2+y^2}\frac{L_{t,V}+L_{r,V}-2y}{L_{t,V}L_{r,V}}dy}.$
%\begin{equation}
%\begin{aligned}
%\gamma _{11}=\int_0^{\frac{\left| L_{t,V}-L_{r,V} \right|}{2}}{\frac{1}{D^2+x^2+y^2}\frac{2}{L_{V,\max}}dy},
%\end{aligned}
%\end{equation}
%\begin{equation}
%\begin{aligned}
%\gamma _{12}\!=\!\!\int_{\frac{\left| L_{t,V}-L_{r,V} \right|}{2}}^{\frac{L_{t,V}+L_{r,V}}{2}}{\frac{1}{D^2+x^2+y^2}\frac{L_{t,V}+L_{r,V}-2y}{L_{t,V}L_{r,V}}dy}.
%\end{aligned}
%\end{equation}
For $\gamma _{11}$, we have
\begin{equation}\label{r11}
\begin{aligned}
\gamma_{11}\overset{\left( a \right)}{=}&\frac{2}{L_{V,\max}}\frac{1}{\sqrt{D^2+x^2}}\left. \arctan{\frac{y}{\sqrt{D^2+x^2}}} \right|_{0}^{\frac{\left| L_{t,V}-L_{r,V} \right|}{2}}\\
&=\frac{2}{L_{V,\max}}\frac{1}{\sqrt{D^2+x^2}}\arctan{\frac{\left| L_{t,V}-L_{r,V} \right|}{2\sqrt{D^2+x^2}}},
\end{aligned}
\end{equation}
where $(a)$ follows \eqref{integral2} by letting $a^2=D^2+x^2$. As for $\gamma _{12}$, based on \eqref{integral1} and \eqref{integral2} by letting $a^2=D^2+x^2$, we can compute $\gamma _{12}$ as \eqref{r12}, where $\mu _1=\left( L_{t,V}-L_{r,V} \right) ^2+4D^2$ and $\mu _2=\left( L_{t,V}+L_{r,V} \right) ^2+4D^2$. Then, by substituting \eqref{r11} and \eqref{r12} into \eqref{r1} and applying \eqref{arctan}, we have
\addtocounter{equation}{1}
\begin{equation}\label{r1_closed}
\begin{aligned}
\gamma _1\!=\!\frac{2L_{t,V}L_{r,V}}{D^2+x^2}\!+\!\ln ( \frac{\mu_1+4x^2 }{\mu_2+4x^2}).
\end{aligned}
\end{equation}

Then, based on $\gamma _1$, we can denote $\gamma =\frac{1}{\left( 4\pi \right) ^2}L_{t,H}L_{r,H}\int_x{\gamma _1f\left( x \right) dx}$ as \eqref{r_final}. To compute \eqref{r_final}, we define $T\left( x \right) \triangleq \int{\gamma _1}dx$ and $Q\left( x \right) \triangleq \int{\gamma _1}xdx$. Relying on \eqref{integral2} and \eqref{integral3}, we can derive $T\left( x \right)$ as given in \eqref{T_function}, where
$
\int{\ln ( \mu _1+4x^2 )}dx=\small{\frac{1}{2}}\int{\ln ( \mu _1+( 2x ) ^2 )}d2x
=x\ln ( \mu _1+4x^2 ) -2x+\sqrt{\mu _1}\arctan{\frac{2x}{\sqrt{\mu _1}}}
$. Then, based on \eqref{integral1} and \eqref{integral4}, we can compute $Q\left( x \right)$ as given in \eqref{Q_function}, where
$\int{x\ln ( \mu _1+4x^2 ) dx}=\frac{1}{4}\int{2x\ln [ \mu _1+( 2x ) ^2 ] d2x}=\frac{1}{2}x^2\ln ( \mu _1+4x^2 ) -\frac{1}{2}x^2+\frac{1}{8}\mu _1\ln ( \mu _1+4x^2 ).
$
Thus, by substituting \eqref{r1_closed} into \eqref{r_final} and applying \eqref{T_function} and \eqref{Q_function}, we can compute $\gamma $ as the form shown in \eqref{Numerator}.

Then, we focus on the computation of the denominator $\xi$ in detail. By substituting the coordinates into \eqref{Kernel_Scalar}, we can represent the auto-correlation kernel function $K(\mathbf{r}_t,\mathbf{r}_{t^{\prime}})$ as \eqref{Kernel_expand}, where step $(a)$ is implemented based on the approximation as shown in \eqref{d_approx}. Note that step $(c)$ in \eqref{d_approx} neglects $1/D^4$-related components and applies \eqref{square}. Besides, step $(c)$ in \eqref{d_approx} follows from \eqref{Taylor}. Moreover, in step $(b)$ of
\eqref{Kernel_expand}, we assume to neglect the effects of $\exp(\cdot)$-related components on the computation of the integral. Then,
by substituting \eqref{Kernel_expand} into ${\int_{S_T}{\int_{S_T}{\left| K(\mathbf{r}_t,\mathbf{r}_{t^{\prime}}) \right|^2}d}\mathbf{r}_td\mathbf{r}_{t^{\prime}}}$, we can represent the denominator of \eqref{CAP_EDoF_Scalar} as \eqref{K_dis}. Note that it is very challenging to directly compute ${\int_{S_T}{\int_{S_T}{\left| K(\mathbf{r}_t,\mathbf{r}_{t^{\prime}}) \right|^2}d}\mathbf{r}_td\mathbf{r}_{t^{\prime}}}$ with the effect of $\exp(\cdot)$-related components due to the strong mutual relationship between each integrated functions. Thus, we impose an assumption to extract the $\exp(\cdot)$-related components and approximately compute them separately from other components as \eqref{K_dis}, where $\varphi =\frac{1}{N_{s}^{2}M_{s}^{2}}\sum_{o=1}^{M_s}{\sum_{u=1}^{M_s}{| \sum_{k^{\prime}=1}^{N_s}{\exp ( j\kappa _0d_{k^{\prime}o} ) \exp ( -j\kappa _0d_{k^{\prime}u}^{\prime} )} |^2}}$ is the approximated $\exp(\cdot)$-related component. Besides,
$d_{k^{\prime}o}=\left| \mathbf{r}_{r,k^{\prime}}-\mathbf{r}_{t,o} \right|$ and $d_{k^{\prime}u}^{\prime}=\left| \mathbf{r}_{r,k^{\prime}}-\mathbf{r}_{t^{\prime},u} \right|$ are the distances between $k^{\prime}$-th uniformly random sampled receiving point and the $o$-th or $u$-th uniformly random sampled transmitting point, with $\mathbf{r}_{r,k^{\prime}}\in S_R$, $\left\{ \mathbf{r}_{t,o},\mathbf{r}_{t^{\prime},u} \right\} \in S_T$, $\left\{ o,u \right\} =\left\{ 1,\dots ,M_s \right\} $, and $k^{\prime}=\left\{ 1,\dots ,N_s \right\}$.

Furthermore, we focus on the computation of $\zeta$ in \eqref{K_dis}.
\addtocounter{equation}{3}
%\begin{equation}
%\zeta =\int_{S_T}{\int_{S_T}{\left| \frac{L_{r,H}L_{r,V}}{\left( 4\pi \right) ^2[ D^2+( \frac{y_t+y_{t^{\prime}}}{2} ) ^2+( \frac{x_t+x_{t^{\prime}}}{2} ) ^2 ]} \right|^2}d}\mathbf{r}_td\mathbf{r}_{t^{\prime}}.
%\end{equation}
Inspired by \cite{2023arXiv230406141X}, we let $y\triangleq \frac{\left| y_t+y_{t^{\prime}} \right|}{2}$ and $x\triangleq \frac{\left| x_t+x_{t^{\prime}} \right|}{2}$, and thus, we can denote $\zeta$ as
\addtocounter{equation}{1}
\begin{equation}\label{F0}
\zeta =\int_x{\int_y{\frac{\mu _3j\left( x \right) w\left( y \right) S_xS_y}{\left( D^2+x^2+y^2 \right) ^2}}}dxdy,
\end{equation}
where $S_x=L_{t,H}^{2}$ is the area of the integration region of $x$, $S_y=L_{t,V}^{2}$ is the area of the integration region of $y$, $j(x)$ is the PDF of $x$, $w(y)$ is the PDF of $y$, and $
\mu _3=\mu _0L_{r,H}^{2}L_{r,V}^{2}$.
Based on the similar method above, we can derive $j(x)$ and $w(y)$ as $j\left( x \right) =\frac{4L_{t,H}-8x}{L_{t,H}^{2}},0\leqslant x\leqslant \frac{L_{t,H}}{2}$ and $w\left( y \right) =\frac{4L_{t,V}-8y}{L_{t,V}^{2}},0\leqslant y\leqslant \frac{L_{t,V}}{2}$, respectively.
%\begin{equation}\label{fy2}
%\left\{ \begin{array}{c}
%	j\left( x \right) =\frac{4L_{t,H}-8x}{L_{t,H}^{2}},0\leqslant x\leqslant \frac{L_{t,H}}{2}\\
%	w\left( y \right) =\frac{4L_{t,V}-8y}{L_{t,V}^{2}},0\leqslant y\leqslant \frac{L_{t,V}}{2}.\\
%\end{array} \right.
%\end{equation}
By substituting $j(x)$ and $w(y)$ into \eqref{F0}, we can compute $\zeta =\int_0^{\frac{L_{t,H}}{2}}{\mu _3\left( 4L_{t,H}-8x \right) \underset{g}{\underbrace{{L_{t,V}^{2}}/{\left( D^2+x^2 \right) ^2}}}}dx \overset{\left( a \right)}{=}\frac{4\mu _3L_{t,H}^{2}L_{t,V}^{2}}{D^2( 4D^2+L_{t,H}^{2} )}+\frac{2\mu _3L_{t,H}L_{t,V}^{2}}{D^3}\arctan{\frac{L_{t,H}}{2D}}
+\frac{4\mu _3L_{t,V}^{2}}{( 4D^2+L_{t,H}^{2} )}-\frac{4\mu _3L_{t,V}^{2}}{D^2}$,
where step $(a)$ is implemented based on \eqref{integral6} and \eqref{integral7}. As for $g$, we have $g=\int_0^{\frac{L_{t,V}}{2}}{\left( 4L_{t,V}-8y \right) \frac{1}{\left( D^2+x^2+y^2 \right) ^2}dy}$ as
\begin{equation}\label{gfunc}
\begin{aligned}
&g%={\int_0^{\frac{L_{t,V}}{2}}{\frac{4L_{t,V}}{\left( D^2+x^2+y^2 \right) ^2}dy}}-{\int_0^{\frac{L_{t,V}}{2}}{\frac{8y}{\left( D^2+x^2+y^2 \right) ^2}dy}}\\
\overset{\left( b \right)}{\approx}\left. \left[ \frac{4L_{t,V}y}{2\left( D^2+x^2 \right) \left( y^2+D^2+x^2 \right)}+\frac{4L_{t,V}y}{2\left( D^2+x^2 \right) ^2} \right] \right|_{0}^{\frac{L_{t,V}}{2}}\\
&+\left. \frac{4}{\left( D^2+x^2+y^2 \right)} \right|_{0}^{\frac{L_{t,V}}{2}}=\frac{L_{t,V}^{2}}{\left( D^2+x^2 \right) ^2}
\end{aligned}
\end{equation}
where step $(b)$ follows from \eqref{integral6}, \eqref{integral7} and applies \eqref{arctan} to approximate $\mathrm{arc}(\cdot)$-related component.

In summary, we can approximately compute \eqref{K_dis} in novel closed-form as \eqref{denominator_close} and complete the proof of Theorem~\ref{CAPclosed}.

\end{appendices}

\bibliographystyle{IEEEtran}
\bibliography{IEEEabrv,Ref}

\begin{thebibliography}{10}
\providecommand{\url}[1]{#1}
\csname url@samestyle\endcsname
\providecommand{\newblock}{\relax}
\providecommand{\bibinfo}[2]{#2}
\providecommand{\BIBentrySTDinterwordspacing}{\spaceskip=0pt\relax}
\providecommand{\BIBentryALTinterwordstretchfactor}{4}
\providecommand{\BIBentryALTinterwordspacing}{\spaceskip=\fontdimen2\font plus
\BIBentryALTinterwordstretchfactor\fontdimen3\font minus
  \fontdimen4\font\relax}
\providecommand{\BIBforeignlanguage}[2]{{%
\expandafter\ifx\csname l@#1\endcsname\relax
\typeout{** WARNING: IEEEtran.bst: No hyphenation pattern has been}%
\typeout{** loaded for the language `#1'. Using the pattern for}%
\typeout{** the default language instead.}%
\else
\language=\csname l@#1\endcsname
\fi
#2}}
\providecommand{\BIBdecl}{\relax}
\BIBdecl

\bibitem{you2021towards}
X.~You, C.-X. Wang, J.~Huang, X.~Gao, Z.~Zhang, M.~Wang, Y.~Huang, C.~Zhang,
  Y.~Jiang, J.~Wang \emph{et~al.}, ``Towards {6G} wireless communication
  networks: Vision, enabling technologies, and new paradigm shifts,''
  \emph{Sci. China Inf. Sci.}, vol.~64, pp. 1--74, Jan. 2021.

\bibitem{9390169}
H.~Tataria, M.~Shafi, A.~F. Molisch, M.~Dohler, H.~Sj{\"o}land, and
  F.~Tufvesson, ``{6G} wireless systems: Vision, requirements, challenges,
  insights, and opportunities,'' \emph{Proc. IEEE}, vol. 109, no.~7, pp.
  1166--1199, Jul. 2021.

\bibitem{9113273}
J.~Zhang, E.~Bj{\"o}rnson, M.~Matthaiou, D.~W.~K. Ng, H.~Yang, and D.~J. Love,
  ``Prospective multiple antenna technologies for beyond {5G},'' \emph{IEEE J.
  Sel. Areas Commun.}, vol.~38, no.~8, pp. 1637--1660, Jun. 2020.

\bibitem{ZheSurvey}
Z.~{Wang}, J.~{Zhang}, H.~{Du}, D.~{Niyato}, S.~{Cui}, B.~{Ai}, M.~{Debbah},
  K.~B. {Letaief}, and H.~V. {Poor}, ``A tutorial on extremely large-scale
  {MIMO} for {6G}: Fundamentals, signal processing, and applications,''
  \emph{IEEE Commun. Surveys Tuts.}, vol.~26, no.~3, pp. 1560--1605, 3rd
  quarter, 2024.

\bibitem{2023arXiv231011044L}
H.~{Lu}, Y.~{Zeng}, C.~{You}, Y.~{Han}, J.~{Zhang}, Z.~{Wang}, Z.~{Dong},
  S.~{Jin}, C.-X. {Wang}, T.~{Jiang}, X.~{You}, and R.~{Zhang}, ``A tutorial on
  near-field {XL-MIMO} communications towards {6G},'' \emph{IEEE Commun.
  Surveys Tuts.}, vol.~26, no.~4, pp. 2213--2257, 4th quarter, 2024.

\bibitem{9903389}
M.~Cui, Z.~Wu, Y.~Lu, X.~Wei, and L.~Dai, ``Near-field communications for {6G}:
  Fundamentals, challenges, potentials, and future directions,'' \emph{IEEE
  Commun. Mag.}, vol.~61, no.~1, pp. 40--46, Jan. 2023.

\bibitem{2022arXiv221201257G}
T.~{Gong}, I.~{Vinieratou}, R.~{Ji}, C.~{Huang}, G.~C. {Alexandropoulos},
  L.~{Wei}, Z.~{Zhang}, M.~{Debbah}, H.~V. {Poor}, and C.~{Yuen},
  ``{Holographic {MIMO} communications: Theoretical foundations, enabling
  technologies, and future directions},'' \emph{IEEE Commun. Surveys Tuts.},
  vol.~26, no.~1, pp. 196--257, 1st quarter, 2024.

\bibitem{8811733}
Q.~Wu and R.~Zhang, ``Intelligent reflecting surface enhanced wireless network
  via joint active and passive beamforming,'' \emph{IEEE Trans. Wireless
  Commun.}, vol.~18, no.~11, pp. 5394--5409, Nov. 2019.

\bibitem{enyusurvey}
E.~{Shi}, J.~{Zhang}, H.~{Du}, B.~{Ai}, C.~{Yuen}, D.~{Niyato}, K.~B.
  {Letaief}, {Xuemin}, and {Shen}, ``{RIS}-aided cell-free massive {MIMO}
  systems for {6G}: Fundamentals, system design, and applications,''
  \emph{Proc. IEEE}, vol. 112, no.~4, pp. 331--364, Apr. 2024.

\bibitem{9808307}
Z.~Xie, W.~Yi, X.~Wu, Y.~Liu, and A.~Nallanathan, ``{STAR-RIS} aided {NOMA} in
  multicell networks: A general analytical framework with gamma distributed
  channel modeling,'' \emph{IEEE Trans. Commun.}, vol.~70, no.~8, pp.
  5629--5644, Aug. 2022.

\bibitem{sang2023multi}
J.~Sang, M.~Zhou, J.~Lan, B.~Gao, W.~Tang, X.~Li, S.~Jin, E.~Basar, C.~Li,
  Q.~Cheng \emph{et~al.}, ``Multi-scenario broadband channel measurement and
  modeling for sub-6 {GHz} {RIS}-assisted wireless communication systems,''
  \emph{IEEE Trans. Wireless Commun.}, to appear, 2023.

\bibitem{du2023ai}
H.~Du, J.~Wang, D.~Niyato, J.~Kang, Z.~Xiong, and D.~I. Kim, ``{AI}-generated
  incentive mechanism and full-duplex semantic communications for information
  sharing,'' \emph{IEEE J. Sel. Areas Commun.}, vol.~41, no.~9, pp. 2981--2997,
  Sep. 2023.

\bibitem{10051719}
X.~Hou, J.~Wang, C.~Jiang, X.~Zhang, Y.~Ren, and M.~Debbah, ``{UAV}-enabled
  covert federated learning,'' \emph{IEEE Trans. Wireless Commun.}, vol.~22,
  no.~10, pp. 6793--6809, Oct. 2023.

\bibitem{[162]}
E.~Bj{\"o}rnson and L.~Sanguinetti, ``Making cell-free massive {MIMO}
  competitive with {MMSE} processing and centralized implementation,''
  \emph{IEEE Trans. Wireless Commun.}, vol.~19, no.~1, pp. 77--90, Jan. 2019.

\bibitem{sanguinetti2019toward}
L.~Sanguinetti, E.~Bj{\"o}rnson, and J.~Hoydis, ``Toward massive {MIMO} 2.0:
  Understanding spatial correlation, interference suppression, and pilot
  contamination,'' \emph{IEEE Trans. Commun.}, vol.~68, no.~1, pp. 232--257,
  Jan. 2020.

\bibitem{OBETrans}
Z.~Wang, J.~Zhang, E.~Bj{\"o}rnson, D.~Niyato, and B.~Ai, ``{Optimal Bilinear
  Equalizer for Cell-Free Massive {MIMO} Systems over Correlated {R}ician
  Channels},'' \emph{arXiv:2407.18531}, 2024.

\bibitem{[15]}
R.~Deng, B.~Di, H.~Zhang, and L.~Song, ``{HDMA}: {Holographic}-pattern division
  multiple access,'' \emph{IEEE J. Sel. Areas Comm.}, vol.~40, no.~4, pp.
  1317--1332, Apr. 2022.

\bibitem{[97]}
L.~{Wei}, C.~{Huang}, G.~C. {Alexandropoulos}, Z.~{Yang}, J.~{Yang}, W.~E.~I.
  {Sha}, Z.~{Zhang}, M.~{Debbah}, and C.~{Yuen}, ``Tri-polarized holographic
  {MIMO} surfaces for near-field communications: Channel modeling and precoding
  design,'' \emph{IEEE Trans. Wireless Commun.}, vol.~22, no.~12, pp.
  8828--8842, Dec. 2023.

\bibitem{nam2013full}
Y.-H. Nam, B.~L. Ng, K.~Sayana, Y.~Li, J.~Zhang, Y.~Kim, and J.~Lee,
  ``Full-dimension {MIMO} ({FD-MIMO}) for next generation cellular
  technology,'' \emph{IEEE Commun. Mag.}, vol.~51, no.~6, pp. 172--179, Jun.
  2013.

\bibitem{kim2014full}
Y.~Kim, H.~Ji, J.~Lee, Y.-H. Nam, B.~L. Ng, I.~Tzanidis, Y.~Li, and J.~Zhang,
  ``Full dimension {MIMO} ({FD-MIMO}): The next evolution of {MIMO} in {LTE}
  systems,'' \emph{IEEE Wireless Commun.}, vol.~21, no.~2, pp. 26--33, Apr.
  2014.

\bibitem{[70]}
Z.~Zhang and L.~Dai, ``Pattern-division multiplexing for multi-user
  continuous-aperture {MIMO},'' \emph{IEEE J. Sel. Areas Commun.}, vol.~41,
  no.~8, pp. 2350--2366, Aug. 2023.

\bibitem{2024arXiv240105900L}
Y.~{Liu}, C.~{Ouyang}, Z.~{Wang}, J.~{Xu}, X.~{Mu}, and A.~L. {Swindlehurst},
  ``Near-field communications: A comprehensive survey,'' \emph{IEEE Commun.
  Surveys Tuts.}, to appear, 2024.

\bibitem{10220205}
Y.~Liu, Z.~Wang, J.~Xu, C.~Ouyang, X.~Mu, and R.~Schober, ``Near-field
  communications: A tutorial review,'' \emph{IEEE Open J. Commun. Soc.},
  vol.~4, pp. 1999--2049, Aug. 2023.

\bibitem{arnoldus2001representation}
H.~F. Arnoldus, ``Representation of the near-field, middle-field, and far-field
  electromagnetic green¡¯s functions in reciprocal space,'' \emph{JOSA B},
  vol.~18, no.~4, pp. 547--555, 2001.

\bibitem{[41]}
A.~Pizzo, L.~Sanguinetti, and T.~L. Marzetta, ``Fourier plane-wave series
  expansion for holographic {MIMO} communications,'' \emph{IEEE Trans. Wireless
  Commun.}, vol.~21, no.~9, pp. 6890--6905, Mar. 2022.

\bibitem{2022arXiv220903082R}
P.~Ramezani and E.~Bj{\"o}rnson, ``Near-field beamforming and multiplexing
  using extremely large aperture arrays,'' in \emph{Fundamentals of 6G
  Communications and Networking}.\hskip 1em plus 0.5em minus 0.4em\relax
  Springer, 2023, pp. 317--349.

\bibitem{[34]}
H.~Lu and Y.~Zeng, ``Near-field modeling and performance analysis for
  multi-user extremely large-scale {MIMO} communication,'' \emph{IEEE Commun.
  Lett.}, vol.~26, no.~2, pp. 277--281, Nov. 2022.

\bibitem{[6]}
M.~Cui and L.~Dai, ``Channel estimation for extremely large-scale {MIMO}:
  {Far}-field or near-field?'' \emph{IEEE Trans. Commun.}, vol.~70, no.~4, pp.
  2663--2677, Apr. 2022.

\bibitem{[35]}
Y.~Lu and L.~Dai, ``Near-field channel estimation in mixed {LoS/NLoS}
  environments for extremely large-scale {MIMO} systems,'' \emph{IEEE Trans.
  Commun.}, vol.~71, no.~6, pp. 3694--3707, Jun. 2023.

\bibitem{chen-jsac3}
Y.~Chen, Y.~Wang, Z.~Wang, and Z.~Han, ``Angular-distance based channel
  estimation for holographic {MIMO},'' \emph{IEEE J. Sel. Areas Commun.},
  vol.~42, no.~6, pp. 1684--1702, Jun. 2024.

\bibitem{10243590}
Z.~Wu, M.~Cui, and L.~Dai, ``Enabling more users to benefit from near-field
  communications: From linear to circular array,'' \emph{IEEE Trans. Wireless
  Commun.}, vol.~23, no.~4, pp. 3735--3748, Apr. 2024.

\bibitem{[101]}
H.~Zhang, N.~Shlezinger, F.~Guidi, D.~Dardari, and Y.~C. Eldar, ``{6G} wireless
  communications: From far-field beam steering to near-field beam focusing,''
  \emph{IEEE Commun. Mag.}, vol.~61, no.~4, pp. 72--77, Apr. 2023.

\bibitem{2023arXiv230616206Y}
C.~{You}, Y.~{Zhang}, C.~{Wu}, Y.~{Zeng}, B.~{Zheng}, L.~{Chen}, L.~{Dai}, and
  A.~L. {Swindlehurst}, ``Near-field beam management for extremely large-scale
  array communications,'' \emph{arXiv:2306.16206}, 2023.

\bibitem{Zhilongmag}
Z.~{Liu}, J.~{Zhang}, Z.~{Liu}, H.~{Du}, Z.~{Wang}, D.~{Niyato}, M.~{Guizani},
  and B.~{Ai}, ``Cell-free {XL-MIMO} meets multi-agent reinforcement learning:
  Architectures, challenges, and future directions,'' \emph{IEEE Wireless
  Commun.}, vol.~31, no.~4, pp. 155--162, Aug. 2024.

\bibitem{franceschetti2017wave}
M.~Franceschetti, \emph{Wave theory of information}.\hskip 1em plus 0.5em minus
  0.4em\relax Cambridge University Press, 2017.

\bibitem{poon2005degrees}
A.~S. Poon, R.~W. Brodersen, and D.~N. Tse, ``Degrees of freedom in
  multiple-antenna channels: A signal space approach,'' \emph{IEEE Trans. Inf.
  Theory}, vol.~51, no.~2, pp. 523--536, Feb. 2005.

\bibitem{franceschetti2015landau}
M.~Franceschetti, ``On {L}andau’s eigenvalue theorem and information
  cut-sets,'' \emph{IEEE Trans. Inf. Theory}, vol.~61, no.~9, pp. 5042--5051,
  Sep. 2015.

\bibitem{pizzo2022landau}
A.~Pizzo and A.~Lozano, ``On {L}andau’s eigenvalue theorem for line-of-sight
  {MIMO} channels,'' \emph{IEEE Wireless Commun. Lett.}, vol.~11, no.~12, pp.
  2565--2569, Dec. 2022.

\bibitem{ouyangEDoF}
C.~{Ouyang}, Y.~{Liu}, X.~{Zhang}, and L.~{Hanzo}, ``Near-field communications:
  A degree-of-freedom perspective,'' \emph{arXiv:2308.00362}, 2023.

\bibitem{[65]}
S.~S.~A. Yuan, Z.~He, X.~Chen, C.~Huang, and W.~E.~I. Sha, ``Electromagnetic
  effective degree of freedom of an {MIMO} system in free space,'' \emph{IEEE
  Antennas Wireless Propagat. Lett.}, vol.~21, no.~3, pp. 446--450, Mar. 2022.

\bibitem{[29]}
Y.~Jiang and F.~Gao, ``Electromagnetic channel model for near field {MIMO}
  systems in the half space,'' \emph{IEEE Commun. Lett.}, vol.~27, no.~2, pp.
  706--710, Feb. 2023.

\bibitem{2023arXiv230406141X}
Z.~{Xie}, Y.~{Liu}, J.~{Xu}, X.~{Wu}, and A.~{Nallanathan}, ``Performance
  analysis for near-field {MIMO}: Discrete and continuous aperture antennas,''
  \emph{arXiv:2304.06141}, 2023.

\bibitem{EDOFTVT}
Z.~{Wang}, J.~{Zhang}, W.~{Yi}, D.~{Niyato}, H.~{Xiao}, and B.~{Ai},
  ``{Effective Degree of Freedom for Near-Field Plane-Based {XL-MIMO} with
  Tri-Polarization},'' \emph{Fron. Inf. Technol. Electron. Eng.}, vol.~25,
  no.~12, pp. 1723--1731, Dec. 2024.

\bibitem{[50]}
{\"O}.~T. Demir, E.~Bj{\"o}rnson, and L.~Sanguinetti, ``Channel modeling and
  channel estimation for holographic massive {MIMO} with planar arrays,''
  \emph{{IEEE} Wireless Commun. Lett.}, vol.~11, no.~5, pp. 997--1001, May
  2022.

\bibitem{poon2011degree}
A.~S. Poon and N.~David, ``Degree-of-freedom gain from using polarimetric
  antenna elements,'' \emph{IEEE Trans. Inf. Theory}, vol.~57, no.~9, pp.
  5695--5709, Sep. 2011.

\bibitem{4418491}
T.~Muharemovic, A.~Sabharwal, and B.~Aazhang, ``Antenna packing in low-power
  systems: Communication limits and array design,'' \emph{IEEE Trans. Inf.
  Theory}, vol.~54, no.~1, pp. 429--440, Jan. 2008.

\bibitem{1003824}
S.~Verdu, ``Spectral efficiency in the wideband regime,'' \emph{IEEE Trans.
  Inf. Theory}, vol.~48, no.~6, pp. 1319--1343, Jun. 2002.

\bibitem{balanis2016antenna}
C.~A. Balanis, \emph{Antenna theory: analysis and design}.\hskip 1em plus 0.5em
  minus 0.4em\relax John Wiley \& Sons, 2016.

\end{thebibliography}

\end{document}